\documentclass[11pt]{amsart}
\usepackage{amssymb}
\usepackage{amsmath,amsthm,amsfonts}  
\usepackage{mathabx}
\usepackage[applemac]{inputenc} 
 \usepackage{graphicx}
\usepackage{pdfsync}
\usepackage{fancyhdr}
\usepackage[T1]{fontenc}

\usepackage{color}
\usepackage{ulem}
\usepackage{cancel}
\usepackage{xcolor}
\usepackage[breaklinks,colorlinks,backref]{hyperref}
\hypersetup{
    colorlinks, 
    linktoc=all, 
    linkcolor=red,  
  citecolor=hyptxt,
  urlcolor=blue}
  \hypersetup{
  citebordercolor=,
  filebordercolor=green,
  linkbordercolor=blue
}
\definecolor{hyptxt}{rgb}{0.7, 0.4, 0.9}
\usepackage{bibtopic}

\newtheorem{prop}{Proposition}[section]
\newtheorem{assum}{Assumption}[section]

\newcommand{\beprop}{\begin{prop}}
\newcommand{\enprop}{\end{prop}}
\newcommand{\bprf}{\begin{proof}} 
\newcommand{\eprf}{\end{proof}\qed}

\renewcommand{\qed}{\hfill $\square$}
\definecolor{hervecolor}{rgb}{0.8,0,0.7}
     
\newcommand{\ket}[1]{|\kern.3ex#1\kern.3ex\rangle}
\newcommand{\bra}[1]{\langle\kern.3ex #1 \kern.3ex|}
\newcommand{\scalar}[2]{\langle\kern.3ex #1 \kern.3ex|\kern.3ex#2\kern.3ex\rangle}

\newcommand{\ii}{\mathsf{i}}

\def\by{\mathbf{y}}
\def\bx{\mathbf{x}}
\def\br{\mathbf{r}}
\def\bq{\mathbf{q}}
\def\bp{\mathbf{p}}
\def\br{\mathbf{r}}
\def\bb{\mathbf{b}}
\def\bQ{\mathbf{Q}}
\def\bP{\mathbf{P}}
\def\bu{\mathbf{u}}
\def\by{\mathbf{y}}

\def\bv{\mathbf{v}}
\def\calH{{\mathcal H}}
\def\H{\mathbb{H}}
\def\R{\mathbb{R}}
\def\N{\mathbb{N}}
\def\C{\mathbb{C}}
\def\Z {\mathbb{Z}}

\def\lg{\langle }
\def\rg{\rangle }

\def\vap{\varpi}

\def\ud{\mathrm{d}}

\def\sfM{\mathsf{M}}

\def\sfMv{\mathsf{M}^{\vap}}
\def\cMv{\mathcal{M}^{\vap}}
\def\sfMva{\mathsf{M}^{\vapa}}
\def\sfMvp{\mathsf{M}^{\vap_{\psi}}}

\def\vapa{\vap_{a\mathcal{W}}}

\def\Om{\Omega}

\def\eal{e^{(\alpha)}}
\def\Lal{L^{(\alpha)}}
\def\cdm{\mathsf{C}_{\mathrm{DM}}}
\def\sI{\mathsf{I}}

\numberwithin{equation}{section}

\title{$2$-D  Covariant Affine Integral Quantization(s)}
\author{Jean Pierre Gazeau, Tomoi Koide,  and Romain Murenzi}
\address{Universit\'e de Paris, Laboratoire APC (UMR 7164), 75205 Paris, France}
\email{gazeau@apc.in2p3.fr}
\address{Instituto de F\'{i}sica, Universidade Federal do Rio de Janeiro, C.P.
68528, 21941-972, Rio de Janeiro, RJ, Brazil}
\email{koide@if.ufrj.br}
\address{The World Academy of Sciences, TWAS/ICTP\\
Via Costiera 1, Trieste 34151, Italy
}
\email{rmurenzi@twas.org}

\date{\today}        

\begin{document}

\maketitle

\begin{abstract}
Covariant affine integral quantization   is studied and applied to the motion of a particle in a punctured plane $\mathbb{R}_{\ast}^2:=\mathbb{R}^2\setminus\{0\}$, for which the phase space is  $\mathbb{R}_{\ast}^2 \times \mathbb{R}^2$. We examine the consequences of different  quantizer operators built from weight functions on  $\mathbb{R}_{\ast}^2 \times \mathbb{R}^2$. To illustrate the procedure, we examine two  examples of weights. The first one corresponds to $2$-D  coherent state families, while the second  one corresponds  to the affine inversion in the punctured plane. The later yields the usual canonical quantization  and a quasi-probability distribution ($2$-D affine Wigner function) which is real, marginal in both position $ \bq$ and momentum $\bp$. 
\end{abstract}



\tableofcontents

\section {Introduction}
\label{intro}
It is known that canonical quantization can meet problems when the phase space associated with the  considered classical system has a non trivial geometry/topology, e.g., periodic geometries \cite{gazsza16}, non-commutative geometries \cite{gazsza11}, or when some impassable boundaries are imposed. In this work we consider an elementary example of this type, namely the phase space for  the motion of a point particle  in the punctured plane $\R^2_\ast:=  \{ \mathbf{r} \in \R^2\, , \, \mathbf{r} \neq \mathbf{0}\}= \mathbb{R}^2\setminus\{0\}$. With this example, we generalise  the content of a previous article 
by two of the present authors \cite{gazmur16}. The later was devoted to the covariant affine integral quantization of the phase space for the motion of a point particle on the open half-line $\R^+_\ast := \{ x \in \R\, , \, x>0\}$. We also extend the content of the more recent \cite{gazkoimur17} devoted to the non-inertial effect on the motion in the rotating plane, for which we applied a particular case of the general procedure exposed here, namely the $2$-D  affine coherent state quantization. 

The fact that the phase space in the $1$-D case  is the open half-plane, and that the later has the group structure of affine transformations of the line, make natural the choice of  quantization respecting this affine  symmetry instead of the Weyl-Heisenberg symmetry.  The latter is based on the translational symmetry of the plane. For a particle moving in the punctured plane, the geometry of the phase space is $\R^2_\ast\times \R^2$ and this manifold can also be equipped with a group structure which makes it isomorphic to the similitude group SIM$(2)$ in two dimensions, hence our interest in exploring the corresponding covariant integral quantization.  As is shown in the present article, and similarly to the $1$D-case, an important issue  of our quantization procedure is the regularisation of the singularity at the origin with the appearance of an extra repulsive, centrifugal-like, potential $\sim 1/r^2$, $r:= \Vert \mathbf{r}\Vert$,  in the quantum version of the kinetic term $p^{2}$, $p:= \Vert \mathbf{p}\Vert$, $\mathbf{p}$ being the momentum of the particle. This term complements  the kinetic operator   $-\Delta$ which would be obtained through the standard (canonical) quantization, namely $ \bq  \mapsto  \bQ $, $ \bp  \mapsto  \bP  $, with $[Q_i,P_j] =\delta_{ij} \ii \hbar \sI$, and $f\left( \bq , \bp \right) \mapsto \mathrm{Sym}f\left( \bQ , \bP  \right)$. Since with our method the strength of such a repulsive potential can be adjusted at will,  essential self-adjointness of the regularized quantum kinetic term can be guaranteed as well, allowing to ignore boundary conditions and to deal with a ``unique physics" \cite{reedsimon2}. On the contrary, a simple application of the  canonical quantization  cannot cure any kind of singularity  present in the classical model, and this drawback is often reflected in the non-essential self-adjointness of some basic observables for the corresponding quantum model. 

Let us present a more explicit discussion about this important point. Classical physical examples of the punctured plane as a configuration space may be found with the motion in a rotating plane (see the recent \cite{gazkoimur17} and references therein) or, of course, in magnetism, where one of the most celebrated illustrations is the Aharonov-Bohm model  \cite{ahabohm59}. The physics of a magnetic flux in an infinitely thin solenoid (called Aharonov-Bohm flux or Aharonov-Bohm vortex) has been investigated both from theoretical and experimental points of view \cite{ahabohm59,olpop85,hamilton87,geystovi04}. To be more precise, let  us consider the motion of a  charged particle, mass $m$, charge $\mathfrak{q}$, confined to a plane and submitted to the action of the singular magnetic field $\mathbf{B}( \br  ) = \Phi\, \delta( \br  )\hat{\mathbf{k}}= \pmb{\nabla} \times \mathbf{A}( \br  )$  where $\mathbf{A}$ is the vector potential and $\hat{\mathbf{k}}$ is the unit vector perpendicular to the plane, $\Phi$ is the magnetic flux through the infinitely thin solenoid, i.e. $\Phi= \int_S \mathbf{B}( \br  )\cdot \pmb{\ud S}= \oint \mathbf{A}(\br)\cdot \ud  \br  $. 
Components of  $\mathbf{A}(\br)$ can be chosen as
\begin{equation}
\label{vecfield}
A_x= -\frac{\Phi}{2\pi} \, \frac{y}{{r}^2}\, , \quad      A_y= \frac{\Phi}{2\pi} \, \frac{x}{{r}^2}\, , \quad  \br   = x \hat{\pmb{\imath}} + y \hat{\pmb{\jmath}} \, ,
\end{equation}
where $(\hat{\pmb{\imath}},\hat{\pmb{\jmath}},\hat{\pmb{k}})$ form a direct orthonormal frame.  
The Lagrangian \cite{lyndnou98} for the motion of the  particle  in such a field is given by 
\begin{equation}
\label{lagr}
\mathcal{L}= \frac{1}{2}m \dot{ \br  }^{\,2} + \mathfrak{q}\, \dot{ \br  }\cdot \mathbf{A}/c\, . 
\end{equation}
The momentum conjugate to $ \br  $ is 
\begin{equation}
\label{cmom}
 \bp  = \frac{\partial \mathcal{L}}{\partial \dot{ \br  }} = m \dot{ \br  }  + \mathfrak{q}\,\mathbf{A}/c\, ,
\end{equation}
which is not a gauge invariant quantity, as it is well known. 
On the other  hand,  the  mechanical momentum of the particle
\begin{equation}
\label{partmom}
m \dot{ \br  } =  \bp    -\mathfrak{q}\,\mathbf{A}/c\, ,
\end{equation}
 is gauge invariant and therefore has greater physical significance.
The Hamiltonian is given by
 \begin{equation}
\label{clham}
H = H(\br,  \bp  ) =  \bp  \cdot \dot\br - \mathcal{L} = \frac{1}{2m} (  \bp   - \mathfrak{q}\,\mathbf{A}/c)^2\,.
\end{equation}
With the expression \eqref{vecfield} at hand, we write this Hamiltonian as the classical observable 
\begin{equation}
\label{exclham}
H( \bq  ,  \bp  ) =  \frac{1}{2m}( \bp  - \mathfrak{q}\,\mathbf{A}/c)^2=  \frac{1}{2m} p^{2}- \frac{\mathfrak{q}}{mc}\,\frac{ \bq  \cdot{ \bp  }}{q^2} + 
\left(\frac{\mathfrak{q}\Phi}{2\pi mc}\right)^2 \, \frac{1}{q^2}\,, \quad q= \Vert \bq\Vert\,,\ p= \Vert \bp\Vert\,, 
\end{equation}
where we have replaced $ \br$  with $ \bq$, a notation more common for phase space variables. 
Due to the presence of the  infinitely thin solenoid, the origin of the plane has to be considered as a singularity, and the classical phase space 
is  $\mathbb{R}_{\ast}^2 \times \mathbb{R}^2$ and \underline{not} just $\mathbb{R}^4$. That geometry is precisely that one for the similitude group in two dimensions.

The organisation of the paper is as follows. 

In Section \ref{covintquant} we give a concise presentation of the method of covariant integral quantisation in view of  application to the main content of the paper where is involved the $2$-D affine group.  
In Section \ref{affG} we describe the main features of this group, also named   the similitude group of the plane and denoted by SIM$(2)$,   and its unitary irreducible representation $U$ which is used in this paper for implementing the corresponding integral quantisation.  
In Section \ref{weight} we build  bounded self-adjoint operators from weight functions defined on the phase space, i.e. on SIM$(2)$, through a kind of ``$2$D-affine transform'', and establish a set of interesting properties, like integral kernels, inverse $2$D-affine transform, connection with affine coherent states, symmetry inversion on the plane, etc. 
In Section \ref{weightex} are given two basic examples of weight functions and corresponding operators. 
In Section \ref{genform} we implement the covariant affine integral quantisation from weights by setting general formulas for arbitrary functions (or distributions) on the $2$-D  affine group SIM$(2)$ and by making explicit the basic operators corresponding to position, momentum, dilation in the punctured plane, angular momentum, etc. 
Section \ref{portrait} is devoted to what we name quantum phase space portrait of the operator corresponding to a function on SIM$(2)$, which means a new function on SIM$(2)$ averaging the latter one
through a trace operation. This gives the key for interpreting within a probabilistic approach the covariant integral quantisation in terms of a sort of an initial coarse-graining, encoded by the weight function, on  SIM$(2)$.  
In Section \ref{inversion} we consider the specific integral quantization based on the Inverse Affine Operator and its interesting quantum phase space portraits. 
In the conclusion, Section \ref{conclu}, we mention some applications to quantum mechanics of the results presented here and also  interesting perspectives concerning  the generalisation of our approach  to the $N$-D affine group. 

In Appendix \ref{AwIvap} we derive a useful formula for the Dirac distribution in the plane. 
In Appendix \ref{QaffCSth} our  quantisation method is implemented in the particular case where we use the so-called $2$-D affine coherent states (ACS), completing in this way the technical content of \cite{gazkoimur17}.

\section{Covariant integral quantizations} \label{covintquant}

In this section, we give a short reminder of what we understand by ``covariant integral quantization". More details can be found in previous papers, like in the recent \cite{bergaz18}. To some extent, this procedure is related to the so-called  Berezin-Toeplitz quantization (see for instance  
the reviews \cite{alienglis05,englis16}).

 Lie group representations \cite{barracz77} offer a wide range of
possibilities for implementing integral quantization(s).  Let $G$
be a Lie group with left Haar measure $\mathrm{d}\mu(g)$, and let
$g\mapsto U\left(g\right)$ be a unitary irreducible representation
(UIR) of $G$ in a Hilbert space $\mathcal{H}$. Let us consider a bounded
operator $\mathsf{M}$ on $\mathcal{H}$ and suppose that the
operator defined by the operator-valued integral
\begin{equation}
\label{boundR}
\mathsf{R}:=\int_{G}\mathsf{M}\left(g\right)\mathrm{d}\mu\left(g\right),\
\mathsf{M}\left(g\right):=U\left(g\right)\mathsf{M}\,U\left(g\right)^{\dagger}\,,
\end{equation}
is bounded  in $\calH$. 
From the left invariance of
$\mathrm{d}\mu(g)$ we have
\begin{equation}
\label{comRU}
U\left(g_{0}\right)\mathsf{R}\,U^{\dagger}\left(g_{0}\right)=\int_{G}\mathsf{M}\left(g_{0}g\right)
\mathrm{d}\mu\left(g\right)=\mathsf{R}\,,
\end{equation}
so $\mathsf{R}$ commutes with all operators $U(g)$, $g\in G$. Thus, from
Schur's Lemma, $\mathsf{R}=c_{\mathsf{M}}\sI$, where $c_{\mathsf{M}}$ is a constant which is supposed to be  non zero. We also suppose that there exists  a unit trace non-negative operator ($\sim$ density operator) $\rho_{0}$ such that this constant is given by the convergent integral
 \begin{equation}
\label{cM}
c_{\mathsf{M}}=\int_{G}\mathrm{Tr}\left(\rho_{0}\mathsf{M}\left(g\right)\right)\mathrm{d}\mu\left(g\right)\,.
\end{equation}
Hence, the  family of operators $\mathsf{M}\left(g\right)$ 
provides the resolution of the identity on $\mathcal{H}$,
\begin{equation}
\int_{G}\mathsf{M}\left(g\right)\mathrm{d}\nu\left(g\right)=\sI,\qquad
\mathrm{d}\nu\left(g\right)
:=\frac{\mathrm{d}\mu\left(g\right)}{c_{\mathsf{M}}}\,.\label{eq:resolution}
\end{equation}
and the subsequent quantization of complex-valued functions (or
distributions, if well-defined) on $G$
\begin{equation}
\label{quantgr} f\mapsto A_{f}=\int_{G}\, \mathsf{M}(g)\,
f(g)\,\mathrm{d}\nu(g)\,.
\end{equation}
This operator-valued integral is  understood in the weak sense, i.e.,   as the sesquilinear form, 
\begin{equation}
\label{quantform}
B_f(\phi_1,\phi_2)= \int_{X}\, \lg \phi_1|\mathsf{M}(g)|\phi_2\rg \, f(g)\, \ud\nu(g)\,.
\end{equation}
The form $B_f$ is  assumed to be  defined on a dense subspace of $\calH$.  If $M$ is self-adjoint and if $f$ is real and at least semi-bounded, the Friedrichs extension \cite[Thm. X.23]{reedsimon80} of $B_f$ univocally defines a self-adjoint operator. However, if $f$ is not semi-bounded, there is no natural choice of a self-adjoint operator associated with $B_f$. In this case, we can consider directly the symmetric operator $A_f$ through its action and related domain in the representation Hilbert space, enabling us to obtain a meaningful self-adjoint extension (unique for particular operators).

The linear map \eqref{quantgr}, function $\mapsto$ operator in $\mathcal{H}$, is
covariant in the sense that
\begin{equation}
\label{covarG} U(g)A_{f}U(g)^{\dagger}=A_{\mathfrak{U}_{r}(g)f}\,.
\end{equation}
In the case when $f\in L^{2}(G,\mathrm{d}\mu(g))$, the action
$(\mathfrak{U}_{r}(g)f)(g^{\prime}) :=f(g^{-1}g^{\prime})$ defines the (left) regular
representation of $G$. Thus, the original symmetry, as is exemplified by the group structure and the arbitrariness in choosing any point in the group manifold 
as representing the group identity, is preserved through the above quantisation procedure. 

The other face of integral quantization concerns a consistent analysis of the operator $A_f$, its \textit{semi-classical portrait}.  It  is implemented
through the study of the so-called lower (Lieb) or covariant (Berezin) symbols. Suppose that $\mathsf{M}$ is
a density operator $\mathsf{M}=
\rho$ on $\mathcal{H}$. Then the operators $\rho(g)$ are also
density, and this allows to build a new function $\widecheck{f}(g)$, called lower or covariant symbol,  as
 \begin{equation}
\label{lowsymb} \widecheck{f}(g)\equiv \widecheck{A}_f:=\int_{G}\,
\mathrm{Tr}(\rho(g)\,\rho(g^{\prime}))\,
f(g^{\prime})\mathrm{d}\nu(g^{\prime})\,.
\end{equation}
The map $f\mapsto \widecheck{f}$ is a generalization of the Berezin or
heat kernel transform on $G$ (see \cite{hall06} and references
therein). Choosing for $\mathsf{M}$ a density operator $\rho$ has multiple advantages, peculiarly
in  regard to probabilistic aspects both on classical and quantum levels \cite{gazhell15}. Finally, under reasonable conditions on $\rho$, we expect that the limit $\widecheck f \to f$ holds in a certain sense as some parameters like $\hbar$ and others tend to some ``classical'' value.

\section{The similitude group SIM$(2)$  (or $2$-D affine group) and its representation $U$} 
\label{affG}

\subsection{Phase space and $2$-D affine group}

As seen in Section \ref{intro}, the space $\Gamma\equiv\mathbb{R}_{\ast}^2\times \mathbb{R}^2=\{( \bq , \bp )\, | \, \bq \in \mathbb{R}_{*}^2\,, \, \bp \in\mathbb{R}^2\}$ can be viewed as the phase space of the (time) evolution of a physical quantity such the position of a particle moving in the punctured plane. It can be identified with the similitude group acting on the plane, also called the $2$D affine group, denoted SIM$(2)$. From now on, all considered quantities have no physical dimension.

In the sequel, Cartesian coordinates of a vector $\bx$ in $\mathbb{R}_{\ast}^2$ or $\R^2$ will be denoted as $\bx= x_1 \mathbf{e}_1 + x_2 \mathbf{e}_2$.   

The group SIM$(2)$ is  formed of translations by $ \bb   \in \mathbb{R}^2$, dilations by $a >0$, and rotations by $\theta \in [0,2\pi)$ mod $(2\pi)$.
The action of an element $(a,\theta, \bb  )$
of SIM(2) on  $\mathbb{R}^2$ is given by:
\begin{equation}
(a,\theta, \bb  )  \bx   = a\,\mathrm{R}(\theta) \bx  + \bb  \, ,
\end{equation}
where $\mathrm{R}(\theta)= \begin{pmatrix}
 \cos\theta     & - \sin\theta   \\
  \sin\theta     &   \cos\theta
\end{pmatrix}$ is the rotation matrix in two dimensions acting on the column vector $\bx$.
The composition law reads as,
\begin{equation}
(a,\theta, \bb  )  (a^{\prime},\theta^{\prime}, \bb  ^{\prime})=(aa^{\prime},\theta+\theta^{\prime},a\,\mathrm{R}(\theta) \bb  ^{\prime}+ \bb  )\, , 
\end{equation}
and the inverse is given by
\begin{equation}
(a,\theta, \bb  )^{-1}=\left(\frac{1}{a}, -\theta, -\frac{\mathrm{R}(-\theta) \bb  }{a}\right)\, . 
\end{equation}

Our   phase space notations $( \bq , \bp ) \equiv (a,\theta, \bb  ) $ for the motion of a particle in the punctured plane $\mathbb{R}_{*}^2=\{ \bx   \in \mathbb{R}^2\,  |\,   \bx    \neq \mathbf{0}\} $ are introduced as
\begin{equation}
q=\frac{1}{a}\, , \  \bq  = (q,\theta ) \,, \  \bb  \equiv \bp \, , 
\end{equation} 
where $(q,\theta)$ are polar coordinates for $\bq$.  
\subsubsection*{Vectors as complex numbers}
In the sequel, we consider  vectors $ \bq $ and $ \bp $ as biunivocally associated with the complex numbers $z( \bq )= qe^{\ii \theta}\equiv  \bq $ and $z( \bp )\equiv  \bp $ (same notation) respectively, and $\mathbf{1}\equiv 1$. Thus,  the multiplication of 2 vectors reads as
\begin{equation}
 \bq \, \bq^{\prime}=(q,\theta ) \,(q^\prime,\theta^\prime ) = (qq^{\prime}, \theta + \theta^{\prime}) \, , 
\end{equation}
(in polar coordinates, and understood mod $2\pi$ for the angular part), and for the inverse,
\begin{equation}
{ \bq }^{-1}=\left(\frac{1}{q},-\theta\right ) \, .
\end{equation}
The multiplication law for the SIM$(2)$ group reads:
\begin{equation}
\label{afflaw}
( \bq , \bp )( \bq^{\prime}, \bp^{\prime})=\left( \bq  \bq^{\prime},\frac{ \bp^{\prime}}{ \bq^{\ast}}+ \bp \right) \, , 
\end{equation}
where $ \bq  ^{\,\ast}=(q,-\theta ) \equiv \bar z( \bq )$ and the inverse,
\begin{equation}
\label{affiinv}
( \bq , \bp )^{-1}=\left(\bq^{-1},- \bq  ^{\,\ast} \bp \right)\,. 
\end{equation}
Hence, SIM$(2)$ is the semidirect product  of the two abelian groups $\R^+_\ast\times \mathrm{SO}(2)$ and $\R^2$:
\begin{equation}
\label{semidir}
\mathrm{SIM}(2)=  \left(\R^+_\ast\times \mathrm{SO}(2)\right)\ltimes\R^2 \sim \C_{\ast} \ltimes \C \,.
\end{equation}
The left action of $\mathrm{SIM}(2)$ on itself is derived from \eqref{affiinv} as 
\begin{equation}
\label{leftaction}
\mathrm{SIM}(2) \ni ( \bq _0, \bp _0): \, ( \bq , \bp ) \mapsto ( \bq^{\prime}, \bp^{\prime})=( \bq _0, \bp _0)( \bq , \bp )=
\left( \bq _0 \bq ,\frac{\bp}{\bq_0^{\ast}}+  \bp \right)\, . 
\end{equation}
The group $\mathrm{SIM}(2)$ is not unimodular. The left invariant Haar measure with respect to \eqref{leftaction} is given by:
\begin{equation}
\ud^2{ \bq }\,\ud^2{ \bp }\, . 
\end{equation}
 This is easily  proved from $\ud^2 \bq^{\prime}=q_0^{2}\ud^2 \bq $ and $\ud^2 \bp^{\prime}=\dfrac{\ud^2 \bp }{q_0}$, and both $\ud^2 \bq $ and $\ud^2 \bp $ are rotationally invariant.  
We also notice that $\dfrac{\ud^2 \bq }{q^2}$ is invariant under complex multiplication and inversion. 

\subsection{$2$-D affine UIR}
The $2$-D affine group SIM$(2)$ has one unitary irreducible representation $U$. It is square integrable and this
is the rationale behind \textit{continuous $2$-D wavelet analysis} \cite{murenzi90,antoine_murenzi08,aagbook13} in two dimensions. This representation is realized
in the Hilbert space
$\mathcal{H}=L^{2}(\mathbb{R}^{2}_\ast,\mathrm{d}^2 \bx  )$ as 
\begin{equation}
(U( \bq , \bp )\psi)( \bx  )=\frac{e^{\ii  \bp \cdot \bx  }}{q}\psi\left(\frac{ \bx  }{ \bq }\right)\,.\label{affrep+}
\end{equation}

Despite that $\mathcal{H}=L^{2}(\mathbb{R}^{2}_\ast,\mathrm{d}^2 \bx  )$ and $L^{2}(\mathbb{R}^{2},\mathrm{d}^2 \bx  )$ are indistinguishable, we will preferably use in the sequel the notation $\int_{\mathbb{R}_{\ast}^2}\ud \bx$ in place of $\int_{\mathbb{R}^2}\ud \bx$ when the integration variable lies in the punctured plane. 

The most immediate (and well-known) orthonormal basis of $\mathcal{H}$ is that one which is  built from Laguerre polynomials and  trigonometric functions,
\begin{align}
\label{LagOB}
 e^{(\alpha)}_{nm}( \bx) &:= \sqrt{\frac{n!}{2\pi (n+\alpha)!}}\, e^{-\frac{x}{2}}\, {x}^{\frac{\alpha-1}{2}}\, L_n^{(\alpha)}(x) e^{\ii m\varphi}\, , \\  \label{LagOB1} &\int_{\mathbb{R}_{\ast}^2}\eal_{nm}( \bx)\, \eal_{n^{\prime}{m^{\prime}}}( \bx) \ud^2 \bx = \delta_{n n^{\prime}}\delta_{m m^{\prime}}\,, \quad n\in \N\, , \quad m\in \Z\,,\quad x=\Vert  \bx   \Vert\, , 
\end{align}
where $\alpha > -1$ is a free parameter, and $(n+\alpha)! = \Gamma(n+\alpha + 1)$. Actually,  since we wish to deal with  functions which, with a certain number of their derivatives,  vanish at the origin, the parameter $\alpha$ should  be imposed to be larger than  some  $\alpha_0 >0$. The matrix elements  $U^{(\alpha)}_{nm,n^{\prime}m^{\prime}}  ( \bq , \bp ):=\lg \eal_{nm}|U( \bq , \bp )|\eal_{n^{\prime}m^{\prime}}\rg$ of the representation $U$ with respect to this basis have the elaborate expression  given, for $m^{\prime}\geq m$, in terms of the integral involving Laguerre Polynomials and Bessel functions
\begin{align}
\label{intform2d} U^{(\alpha)}_{nm,n^{\prime}m^{\prime}}  ( \bq , \bp )  &= \frac{\ii^{m-m^{\prime}}\,e^{-\ii  m^{\prime}\theta}\, e^{-\ii (m- m^{\prime})\psi}}{q^{(\alpha+1)/2}}\, \sqrt{\frac{n!n^{\prime}!}{(n+\alpha)!(n^{\prime}+\alpha)!}} \times\\
\nonumber &\times \int_{0}^{\infty} \ud x\, e^{-\left(\frac{1}{2} + \frac{1}{2q}\right)x}\, x^{\alpha}\, \Lal_n(x)\, \Lal_{n^{\prime}}\left(\frac{x}{q}\right)\, J_{m^{\prime}-m}(-px)\,,
\end{align}
where $ \bp =(p,\psi ) $. Note the relations $J_{m^{\prime}-m}(-px)= (-1)^{m^{\prime}-m}J_{m^{\prime}-m}(px)$ and $J_{m^{\prime}-m}(-px) = (-1)^{m^{\prime}-m}J_{m-m^{\prime}}(-px)$ if $m^{\prime}-m<0$.

One easily verifies the unitarity
\begin{equation}
\label{unitarity}
U^{(\alpha)}_{nm,n^{\prime}m^{\prime}}\left( \bq^{-1},- \bq^{\ast} \bp \right)= \left({U^{(\alpha)}}^{\dag}\right)_{{nm,n^{\prime}m^{\prime}}}( \bq , \bp )=  \overline{U^{(\alpha)}_{{n^{\prime}m^{\prime}},nm}( \bq , \bp )}\,. 
\end{equation}

These matrix elements obey orthogonality relations for the SIM$(2)$ group. Since this group is not unimodular, there exists a positive self-adjoint and invertible operator 
$\mathsf{C}_{\mathrm{DM}}$, named Duflo-Moore operator, such that \cite{aagbook13} the relation
\begin{equation}
\label{orthaffine}
\int_{\Gamma} \mathrm{d}^2 \bq \,\mathrm{d}^2 \bp  \lg U( \bq , \bp )\psi_1| \phi_1\rg \, \overline{\lg U( \bq , \bp )\psi_2| \phi_2\rg}= \langle \mathsf{C}_{\mathrm{DM}}\psi_1|\mathsf{C}_{\mathrm{DM}}\psi_2\rangle \langle \phi_2|\phi_1\rangle\, , 
\end{equation}
holds for any $\psi_1$, $\psi_2$, $\phi_1$, and $\phi_2$ in $\mathcal{H}$.

It is defined as the multiplication operator
\begin{equation}
\label{CDM}
\mathsf{C}_{\mathrm{DM}}\psi( \bx  ) := \dfrac{2\pi}{x}\psi( \bx  )\equiv \dfrac{2\pi}{Q}\psi( \bx  )\,,
\end{equation}
where $ \bQ \psi( \bx  ) :=  \bx   \psi( \bx  )$, with $Q\psi( \bx  ) =  x \psi( \bx  )$,  is the basic position operator with the $| \bx  \rg= \delta_{ \bx  }$'s as eigendistributions.  Thus, the admissibility condition for $\psi \in \mathcal{H}$ amounts to
 \begin{equation}
\label{admcond}
\Vert \mathsf{C}_{\mathrm{DM}}\psi\Vert^2=(2 \pi)^2 \int_{\mathbb{R}_{\ast}^{2}}\frac{\ud^2  \bx  }{x^2} \vert \psi( \bx  )\vert^2 < \infty\, ,
\end{equation}
i.e., 
\begin{equation}
\label{domCDM}
\psi \in L^{2}(\mathbb{R}_{\ast}^2,\mathrm{d}^2 \bx  ) \cap L^{2}\left(\mathbb{R}_{\ast}^{2},\frac{{\mathrm{d}^2 \bx  }}{x^2}\right)\, , 
\end{equation}
which implies that $\underset{ \bx  \to 0^+}{\lim} \psi( \bx  ) = 0$.
For the sequel, due to the non-unimodular nature of SIM$(2)$, that operator $\cdm$ is  expected to play an important r\^ole. A first important and direct consequence of Eq\;\eqref{orthaffine} is the resolution  of the identity in $\mathcal{H}$:
\begin{equation}
\label{residCDM}
\frac{1}{\Vert \mathsf{C}_{\mathrm{DM}}\psi\Vert^2}\int_{\Gamma} \mathrm{d} \bq \,\mathrm{d} \bp  \, U( \bq , \bp )|\psi\rg\lg \psi|U^{\dag} ( \bq , \bp )= \sI\, . 
\end{equation}
satisfied by the family  of \textit{$2$-D affine coherent states} (ACS) 
\begin{equation}
\label{affcS2d}
U( \bq , \bp )|\psi\rg \equiv |\psi_{ \bq , \bp }\rg \,. 
\end{equation}
built from  the  admissible vector $\psi$ through the unitary affine transport $U$. 
 
 Note that the equation \eqref{orthaffine} implies for the matrix elements the integral formula,
\begin{equation}
\label{intmatel}
\int_{\Gamma} \mathrm{d}^2 \bq \,\mathrm{d}^2 \bp \, \overline{U^{(\alpha)}_{n_1m_1,n_1^{\prime}m_1^{\prime}}( \bq , \bp )}\, U^{(\alpha)}_{n_2,m_2,n_2^{\prime}m_2^{\prime}}( \bq , \bp )= 4\pi^2\, \delta_{n_1n_2}\,\delta_{m_1m_2}\,
\lg  \eal_{n_1^{\prime}m_1^{\prime}}|(1/Q^2)\eal_{n_2^{\prime}m_2^{\prime}}\rg\, ,
\end{equation}
which is valid for all $\alpha >0$.

Finally for a trace class operator A, we can use indifferently the 2D affine coherent state decomposition of the identity \eqref{residCDM} or the basis\eqref{LagOB} for the computation
\begin{align}
\label{trace1} \mathrm{Tr}\left(A\right) & = \frac{1}{\Vert \mathsf{C}_{\mathrm{DM}}\psi\Vert^2}\int_{\Gamma} \mathrm{d}^2 \bq^{\prime}\,\mathrm{d}^2 \bp^{\prime} \lg U( \bq^{\prime}, \bp^{\prime})
 \psi|\, A\,U ( \bq^{\prime}, \bp^{\prime})\,|\psi\rg  \\ 
\label{trace2}  & =\sum_{n=0}^{+\infty}\sum_{m=-\infty}^{+\infty} \lg e^{(\alpha)}_{nm}|\, Ae^{(\alpha)}_{nm}\rg   \, .
\end{align}

\section{Bounded self-adjoint operators from weight functions}
\label{weight}

A general method to get density or more general (symmetric) bounded operators consists in proceeding with the following ``SIM$(2)$ transform''.  
Let us choose like in \cite{balfrega14} (see also \cite{bergaz14,becugaro17_1,gazeau_intsympro18} for the Weyl-Heisenberg group,  and \cite{gazeau_olmo19} for the unit disk) a  (weight) function $\mathsf{\varpi}( \bq , \bp )$ on the phase space $\Gamma$
such that the integral 
\begin{equation}
\label{affbop}
\int_{\Gamma}\cdm^{-1} U( \bq , \bp)\cdm^{-1} \,\varpi( \bq , \bp )\,\mathrm{d} \bq \,\mathrm{d} \bp :=\sfMv \, ,
\end{equation}
exists in the weak sense.  In this article, we use the qualifier ``weight'' in a loose sense because $\mathsf{\varpi}$ might assume negative values. The non-unimodularity of SIM$(2)$ justifies the double presence of the inverse of the Duflo-Moore operator, at the difference of the situation one meets with the Weyl-Heisenberg group in \cite{bergaz14}. Note that  in the case of the Heisenberg group, the operator $\sfMv$  is referred as the Fourier transform of $\vap$ on the group, and this transform is invertible under some conditions \cite{daubgross80,daubgrossreig83}. For other groups like semi-direct products see for instance \cite{kumahara76,hiraischiff82}. In our case, the group manifold and the phase space coincide. Moreover, we will need throughout the sequel extra assumptions like positiveness, finiteness, differentiability up to a certain order,  of the functions defined by the integrals
\begin{equation}
\label{Omu}
\Omega_{\beta}( \bu):= \int_{\mathbb{R}_{\ast}^2}\frac{\ud^2  \bq }{q^{\beta +2}}\,\widehat{\vap}_p\left( \bu,\pm \bq \right)\,, \quad \Omega( \bu):= \Omega_{0}( \bu)\,  ,
\end{equation}
where $\widehat{\vap}_p$ is the partial $2$-D Fourier transform of $\vap$ with respect to the variable $ \bp $. This transform is defined as
\begin{equation}
\label{parcoure} \widehat{\vap}_p( \bq , \bx  )=
\frac{1}{2\pi}\int_{\mathbb{R}^2} \ud^2 \bp  e^{-\ii  \bp \cdot \bx  } \vap( \bq , \bp )\, .
\end{equation}
Hence, to start with a minimal set of reasonable properties of   $\vap$, we impose the following. 
\begin{assum}
\label{assumv}
\begin{enumerate}
  \item[(i)] The  function $\vap( \bq , \bp )$ is $C^{\infty}$ on $\Gamma$.
  \item[ii)] It defines a tempered distribution with respect to the variable $ \bp $ for all $ \bq \neq \mathbf{0}$. 
 \item[(iii)]  The operator $\sfMv$is defined by \eqref{affbop} as a  self-adjoint bounded operator on $\calH$.
\end{enumerate}
\end{assum}
We now  establish the nature of $\sfMv$ as an integral operator in $\mathcal{H}$.
 \beprop
 \label{propMomker}
With the assumptions \ref{assumv},  the action of $\sfMv$ on $\phi$ in  $\mathcal{H}$   is given in the form of the linear integral  operator
\begin{equation}
\label{acMom1}
(\sfMv \phi)( \bx  ) = \int_{\R_{\ast}^2}\mathcal{M}^{\vap}( \bx  , \bx^{\prime})\,\phi( \bx^{\prime})\,\ud^2  \bx^{\prime}\, , 
\end{equation}
where the kernel $\cMv$ is given by
\begin{equation}
\label{kerMom}
\mathcal{M}^{\vap}( \bx  , \bx^{\prime}) = \frac{1}{2\pi}\,\frac{ x^2}{ {x^{\prime}}^2}\,\widehat{\vap}_p\left(\frac{ \bx  }{ \bx^{\prime}}, - \bx  \right)\, . 
\end{equation}
\enprop
\bprf It is sufficient to restrict the demonstration  to  the dense linear space $C^{\infty}_0(\R^2_\ast)$ of smooth functions with compact support in the punctured plane.  Let $\phi_1$, $\phi_2$ in $C^{\infty}_0(\R_{\ast}^2)$. We have from  the action on the right of $U( \bq , \bp )$ and of the Duflo-Moore operators
\begin{align*}
&\lg \phi_1 | \sfMv|\phi_2\rg \\
&=\int_{{\R_{\ast}^2}} \ud^2  \bq\int_{{\R^2}} \ud^2  \bp \;\vap( \bq , \bp )\langle \phi_1|\cdm^{-1} U( \bq , \bp )\cdm^{-1}|\phi_2\rangle \\
&=\int_{{\mathbb{R}^{2}_\ast}}\ud^2  \bq\int_{{{\mathbb{R}^{2}}}}  \ud^2  \bp \;\vap( \bq , \bp )\int_{\R_{\ast}^2}\;\ud^2  \bx  \;\overline{\phi_1( \bx  )}\left(\cdm^{-1} U( \bq , \bp )\cdm^{-1}\phi_2\right)( \bx  )\\
&=\int_{\mathbb{R}_\ast^{2}} \ud^2  \bq\int_{{\R^2}} \ud^2  \bp \;\vap( \bq , \bp )\int_{\mathbb{R}_{\ast}^{2}}\;\ud^2  \bx  \;\overline{\phi_1(\bx)}\frac{x}{2\pi}\frac{e^{\ii  \bp \cdot \bx  }}{q}\frac{x}{2\pi q}\phi_2\left(\frac{ \bx  }{ \bq }\right)  \\\
&= \frac{1}{2\pi}\int_{\mathbb{R}_{\ast}^{2}} \frac{\ud^2  \bq }{q^2} \int_{\mathbb{R}_{\ast}^{2}} \ud^2  \bx   \, x^2 \, \overline{\phi_1( \bx  )}\, \phi_2\left(\frac{ \bx  }{ \bq }\right) \, \widehat{\vap}_p( \bq ,- \bx  )
\\
&= \int_{\mathbb{R}_{\ast}^{2}} \ud^2  \bx   \, \overline{\phi_1( \bx  )}\left\{\frac{1}{2\pi}\int_{\mathbb{R}^{2}_{\ast}} \frac{\ud^2  \bq }{q^2} \, x^2\;\phi_2\left(\frac{ \bx  }{ \bq }\right) \, \widehat{\vap}_p( \bq ,- \bx  )\right\}\\\nonumber
&= \int_{\mathbb{R}_{\ast}^{2}}\ud^2  \bx   \,\overline{\phi_1( \bx  )}\left\{\int_{\mathbb{R}_{\ast}^{2}} \ud^2  \bx^{\prime} \,  \, \mathcal{M}^{\vap}( \bx  , \bx^{\prime})\phi_2( \bx^{\prime})\right\}\,,\ \mathcal{M}^{\vap}( \bx  , \bx^{\prime})= \frac{1}{2\pi} \frac{x^2}{{x^{\prime}}^2}\, \widehat{\vap}_p\left(\frac{ \bx  }{ \bx^{\prime}},- \bx  \right)\,,
\end{align*} 
after changing the variable $ \bx  / \bq  \mapsto  \bx^{\prime}$, all permutations of integrals being legitimate due to our assumptions on $\phi_1$, $\phi_2$, and $\vap( \bq , \bp )$.  
\eprf\\

The self-adjointness assumption on $\sfMv$ entails the following symmetry of its integral kernel:
\begin{equation}
\label{symMv}
\cMv( \bx  , \bx^{\prime})= \overline{\cMv( \bx^{\prime}, \bx)}\,, 
\end{equation}
and this allows us to establish a symmetry property on its corresponding  function $\vap$.
\begin{prop}
\label{symvap}
The  function $\varpi( \bq , \bp )$ which defines the bounded self-adjoint $\sfMv$ obeys the following symmetry:
\begin{equation}
\label{eq:condition1}
\varpi( \bq , \bp )=\frac{1}{q^2}\overline{\varpi\left( \bq^{-1},- \bq^{*} \bp \right)}\,.
\end{equation}
\end{prop}
\bprf
Although \eqref{eq:condition1} is a straightforward  consequence of the integral representation \eqref{affbop} of $\sfMv= {\sfMv}^{\dag}$, and the inverse formula \eqref{affiinv}, let us derive it from the symmetry property \eqref{symMv} of the integral kernel given by \eqref{kerMom}. Thus, 
we find that the function $\vap$ must  satisfy 
\begin{equation*}
 \widehat{\vap}_p\left(\frac{ \bx  }{ \bx^{\prime}},- \bx  \right)= \frac{{x^{\prime}}^4}{x^4}\, 
\overline{\widehat{\vap}_p\left(\frac{ \bx^{\prime}}{ \bx  },- \bx^{\prime}\right)}\, , 
\end{equation*}
and so, by a simple change of variables in the integral, 
 \begin{equation*}
 \int_{{\mathbb{R}^{2}}}\ud^2 \bp \;e^{\ii  \bp . \bx  }\vap\left( \bq , \bp \right)=  \int_{{\mathbb{R}^{2}}}\ud^2 \bp \;e^{\ii  \bp . \bx}\frac{1}{q^2}\,\overline{\vap\left( \bq^{-1},- \bq  ^{\,*}\; \bp \right)}\, . 
 \end{equation*}
Hence, the equality \eqref{eq:condition1} results from  inverse  Fourier transform. 
\eprf\\ 

Let us now establish a necessary  condition on $\vap$ to have a unit trace operator $\sfMv$.
\beprop
\label{neccondtr1}
\begin{enumerate}
  \item[(i)] We have the trace formula:
  \begin{equation}
\label{tracecdmU}
 \mathrm{Tr}\left(\cdm^{-1} U( \bq , \bp )\cdm^{-1}\right)= \delta( \bq -1)\delta( \bp )\, ,
\end{equation} 
where $\delta(\bx):= \delta(x_1)\,\delta(x_2)$.
  \item[(ii)] Suppose that the operator $\sfMv$ is unit trace class, $\mathrm{Tr}(\sfMv) = 1$. Then  its corresponding   function $\vap( \bq , \bp )$  obeys
\begin{equation}
\label{Tr1cond}
  \vap(1,\mathbf{0})= 1\, . 
\end{equation}
\end{enumerate}

\enprop
\bprf
From the definition \eqref{affbop} of the operator $\sfMv$ the first step of the proof consists in determining the trace of the operator $\cdm^{-1} U( \bq , \bp )\cdm^{-1}$. For that we use the  \eqref{trace1} with an admissible vector $\psi$. 
We have 
\begin{align*}
\label{}
 \nonumber \mathrm{Tr}\left(\cdm^{-1} U( \bq , \bp )\cdm^{-1}\right)  =& \frac{1}{\Vert \mathsf{C}_{\mathrm{DM}}\psi\Vert^2}\int_{\Gamma} \mathrm{d}^2 \bq^{\prime}\,\mathrm{d}^2 \bp^{\prime}
 \lg \psi|U^{\dag}( \bq^{\prime}, \bp^{\prime})\, \cdm^{-1} U( \bq , \bp )\cdm^{-1}\,U ( \bq^{\prime}, \bp^{\prime})\,|\psi\rg   \\
\nonumber   = &  \frac{1}{(2\pi)^2 q^2\Vert \mathsf{C}_{\mathrm{DM}}\psi\Vert^2}\int_{\mathbb{R}_{\ast}^2}\frac{ \mathrm{d}^2 \bq^{\prime}}{{q^{\prime}}^2}\int_{\mathbb{R}_{\ast}^2}\ud^2  \bx   \, x^2\, e^{\ii  \bp . \bx  }\, \overline{\psi\left(\frac{ \bx  }{ \bq^{\prime}}\right)} \, \psi\left(\frac{ \bx  }{ \bq    \bq^{\prime}}\right)\times\\
\nonumber \times & \int_{\mathbb{R}^{2}}\ud^2  \bp^{\prime}\, e^{-\ii  \bp^{\prime}. \bx  \left(\frac{ \bq -1}{ \bq }\right)}\,. 
\end{align*}
Integrating on $ \bp^{\prime}$ gives $(2\pi)^2 \delta\left( \bx  \,\dfrac{ \bq -1}{ \bq }\right)=  \dfrac{(2\pi)^2 q^2}{x^2}\, \delta( \bq -1)$, which allows to fix $ \bq =1$. Then the change of variable $ \bq^{\prime} \mapsto \by=\dfrac{ \bx  }{ \bq^{\prime}}$ allows to separate the two remaining integrals. That one on 
$\vert\psi(\by)\vert^2/y^2$ yields  $\dfrac{\Vert \mathsf{C}_{\mathrm{DM}}\psi\Vert^2}{(2\pi)^2}$. Finally, 
\begin{equation*}
 \mathrm{Tr}\left(\cdm^{-1} U( \bq , \bp )\cdm^{-1}\right)= \delta( \bq -1)\delta( \bp )\, .
\end{equation*} 
Then, 
$$\mathrm{Tr}(\sfMv)= \int_{\Gamma}\,\mathrm{Tr}\left(\cdm^{-1} U( \bq , \bp )\cdm^{-1}\right)\,\varpi( \bq , \bp )\,\mathrm{d} \bq \,\mathrm{d} \bp  = \vap(1,\mathbf{0})\, . $$
\eprf\\
Note that the condition \eqref{Tr1cond} can be also written as a condition on the function $\Omega_{-2}$.
\begin{equation}
\label{Tr2cond}
\mathrm{Tr}(\sfMv)= \frac{1}{2\pi}\int_{\mathbb{R}_{\ast}^2}\ud^2  \bx   \,\widehat{\vap}_p(1,- \bx  )=  \frac{1}{2\pi} \Omega_{-2}(1)=  1\, .
\end{equation}
 
We now present  the inverse SIM$(2)$  transform which allows to build the weight $\vap$ from the operator
$\sfMv$. 
 \beprop
 \label{propInverseMomker}
Given the operator $\sfMv$ as defined in \eqref{affbop}, and the action of $\sfMv$ on $\phi$ in  $\mathcal{H}$   as given in the form of the  integral  operator
\begin{equation}
\label{acMom2}
(\sfMv \phi)( \bx  ) = \int_{\R_{\ast}^2}\mathcal{M}^{\vap}( \bx  , \bx^{\prime})\,\phi( \bx^{\prime})\,\ud  \bx^{\prime}\, , 
\end{equation}
where the kernel $\cMv$ is given by
\begin{equation}
\mathcal{M}^{\vap}( \bx  , \bx^{\prime}) = \frac{1}{2\pi}\,\frac{ x^2}{ {x^{\prime}}^2}\,\widehat{\vap}_p\left(\frac{ \bx  }{ \bx^{\prime}}, - \bx  \right)\,,
\end{equation}
one   retrieves the function $\vap$ through the following inversion formula:
\begin{equation}
\label{InversionMom}
\vap( \bq , \bp )=\frac{1}{q}\overline{\mathrm{Tr}\left\{U( \bq , \bp )\sfMv\right\}}\, . 
\end{equation}
\enprop
\bprf 
From the definition \eqref{affbop} of $\sfMv$ and the assumption \ref{assumv} one can permute trace and integral  to obtain 
\begin{equation*}
 \mathrm{Tr}\{U( \bq , \bp )\sfMv\}=  \int_{\R_{\ast}^2} \ud^2 \bq^{\prime} \int_{\R^2} \ud^2 \bp^{\prime}\, \mathrm{Tr}\left\{U( \bq , \bp )\,C_{DM}^{-1}\,U( \bq^{\prime} , \bp^{\prime} )\,C_{DM}^{-1} \right\}\, \vap( \bq^{\prime} , \bp^{\prime} )\,.
\end{equation*}
One proves easily the commutation rule 
\begin{equation}
\label{comDM}
\mathrm{Tr}\left\{U( \bq , \bp ) \, C_{DM}^{-1}\right\}=\frac{1}{q} \mathrm{Tr}\left\{C_{DM}^{-1} \,U( \bq , \bp ) \right\}\,.
\end{equation}
Hence, by using the above,  the affine composition law \eqref{afflaw} and the trace formula \eqref{tracecdmU},
\begin{align*}
 \mathrm{Tr}\{U( \bq , \bp )\sfMv\}&= \frac{1}{q} \int_{\R_{\ast}^2} \ud^2 \bq^{\prime} \int_{\R^2} \ud^2 \bp^{\prime}\, \mathrm{Tr}\left\{C_{DM}^{-1}\, U( \bq , \bp )\,U( \bq^{\prime} , \bp^{\prime} )\,C_{DM}^{-1} \right\}\, \vap( \bq^{\prime} , \bp^{\prime} )\\
 \nonumber &= \frac{1}{q} \int_{\R_{\ast}^2} \ud^2 \bq^{\prime} \int_{\R^2} \ud^2 \bp^{\prime}\, \mathrm{Tr}\left\{C_{DM}^{-1}\, U\left( \bq\bq^{\prime} , \frac{\bp^{\prime}}{\bq^{\ast}} +\bp \right)\,C_{DM}^{-1} \right\}\, \vap( \bq^{\prime} , \bp^{\prime} )\\
  \nonumber &=   \frac{1}{q} \int_{\R_{\ast}^2} \ud^2 \bq^{\prime} \int_{\R^2} \ud^2 \bp^{\prime}\,\delta\left( \bq\bq^{\prime}-1\right)\,\delta\left(\frac{\bp^{\prime}}{\bq^{\ast}} +\bp\right)\, \vap( \bq^{\prime} , \bp^{\prime} )\,.
\end{align*}
From the formulas resulting from change of variables in Dirac delta distributions, 
\begin{equation*}
\delta\left( \bq\bq^{\prime}-1\right)= \frac{1}{q^2}\delta\left(\bq^{\prime}-\frac{1}{\bq}\right)\, , \quad \delta\left(\frac{\bp^{\prime}}{\bq^{\ast}} +\bp\right)= q^2\delta\left(\bp^{\prime}+ \bq^{\ast}\bp\right)\, ,
\end{equation*}
we get
\begin{equation}
\label{trUom}
 \mathrm{Tr}\{U( \bq , \bp )\sfMv\}=\frac{1}{q} \, \vap\left( \frac{1}{\bq},  -  \bq^{\ast}\,\bp\right) = q \,\overline{\vap( \bq , \bp )}\,,
\end{equation}
the last equality resulting from \eqref{eq:condition1}.
\eprf\\ 

\section{Examples of weight functions and corresponding operators}
\label{weightex}
We consider here two elementary examples of weights. The first one is  derived from the $2$-D affine coherent states
and the second one is derived from the so-called inversion of operator.

\subsection{Weight related to a pure state}
\label{covcs}
The function $\varpi_{\psi}$ associated to the pure state $|\psi\rg\lg \psi|$  is given by the following proposition, whose proof is straighforward.
\beprop
The operator $\sfMvp$  corresponding to  the function 
\begin{equation}
\label{weightcs}
\varpi_{\psi}( \bq , \bp )=\frac{\langle \psi_{( \bq , \bp )}|\psi \rangle}{q}\, . 
\end{equation}
 is the pure state $|\psi\rg\lg \psi|$ and vice-versa. 
\enprop
 The partial Fourier transform of this function $\vap$ is given  by
\begin{equation}
\label{acsvap}
\widehat{\vap}_{\psi,p}( \bu,\bv)= 2\pi\, \frac{1}{u^2}\, \psi(-\bv)\,\overline{\psi\left(-\frac{\bv}{ \bu}\right)}\,.
\end{equation}

In particular, we have the following relations:
\begin{align}
\label{om1- x}
\widehat{\vap}_{\psi,p}(1,- \bx  )=2\pi \,\vert \psi( \bx  )\vert^2\,, \quad \varpi_{\psi}(1,\mathbf{0})=\Vert\psi\Vert^2\equiv 1\,. 
\end{align}
The weight \eqref{weightcs} satisfies \eqref{eq:condition1} and \eqref{Tr1cond}, as expected. For the former property, we can check:
\begin{align*}
\frac{1}{q^2}\varpi_{\psi}\left( \bq^{-1},- \bq  ^{\,*} \bp \right)
&=\frac{1}{q^2}\frac{\left\langle U\left( \bq^{-1},- \bq  ^{\,*} \bp \right)\psi|\psi \right\rangle}{q^{-1}}=\frac{\langle \psi|U( \bq , \bp )\psi \rangle}{q}
=\frac{\overline{\langle U( \bq , \bp ) \psi|\psi \rangle}}{q}\\\nonumber
&=\overline{\varpi_{\psi}( \bq , \bp )}\,.
\end{align*}

 \subsection{Weight related to the inversion in the punctured plane}
Let us define the unitary inversion operator $\mathcal{I}$ as
\begin{equation}\label{invop}
(\mathcal{I} \phi)( \bx  ):=\frac{1}{x^2}\phi\left(\frac{1}{ \bx  }\right)\, , 
\end{equation}
and the function $\vap_{a\mathcal{W}}$ as yielding the operator
\begin{equation}
\label{MpaW}
\mathsf{M}^{\vap_{a\mathcal{W}}}=2 \mathcal{I}\,,
\end{equation}
 The factor $2$ has been chosen in order to have $\mathrm{Tr} [\mathsf{M}^{\vap_{a\mathcal{W}}}]=1$. 
\beprop 
 The  function $\vap_{a\mathcal{W}}$ is given by:
\begin{equation}
\label{affwigwgt}
\vap_{a\mathcal{W}}( \bq , \bp )=\frac{2}{q}\overline{\mathrm{Tr} [U( \bq , \bp )\mathcal{I}]}=\frac{e^{-\ii\bp \cdot\sqrt{\bq}}}{q}\, , 
\end{equation}
where $\sqrt{ \bq }:=\left(\sqrt{q}, \frac{\theta}{2}\right ) $ if $ \bq =(q,\theta ) $, $0\leq \theta<2\pi$. 
\end{prop}
 
\bprf
Let us prove Equation $\eqref{affwigwgt}$ by using \eqref{InversionMom} and \eqref{residCDM} with an admissible $\psi$. 
\begin{align*}
&\mathrm{Tr}\left(U( \bq , \bp )\mathcal{I}\right)\\
\nonumber
&=\frac{1}{{\Vert C_{DM}\psi\Vert }^2}\int_{\R_{\ast}^2} \ud^2 \bq^{\prime} \int_{\R^2} \ud^2 \bp^{\prime}
\langle \psi_{ \bq^{\prime} \bp^{\prime}}|U( \bq , \bp )\mathcal{I}|\psi_{ \bq^{\prime} \bp^{\prime}}\rangle\\
\nonumber &=\frac{1}{{\Vert C_{DM}\psi\Vert }^2} \int_{\R_{\ast}^2} \ud^2 \bq^{\prime} \int_{\R^2} \ud^2 \bp^{\prime}
\int_{\R_{\ast}^2} \ud^2 \bx   \, \overline{\psi_{ \bq^{\prime} \bp^{\prime}}( \bx  )}\, \left(U( \bq , \bp )\mathcal{I}\psi_{ \bq^{\prime} \bp^{\prime}}\right)( \bx  )
\\\nonumber
&= \frac{(2\pi)^2\,q}{\Vert C_{DM}\psi\Vert^2} \int_{\R_{\ast}^2} \frac{\ud^2 \bq^{\prime}}{q^{{\prime}2}}
\int_{\R_{\ast}^2} \frac{\ud^2 \bx}{x^2} \,\overline{\psi\left(\frac{ \bx  }{ \bq^{\prime}}\right)} \,\psi\left(\frac{ \bq }{ \bq^{\prime} \bx  }\right)\,e^{\ii \bp \cdot \bx  }\,\delta\left( \bx  -\frac{ \bq }{ \bx  }\right)\,,
\end{align*}
where the appearance of the $\delta$ comes from the integration on the variable $ \bp^{\prime}$.
Let us use the following identity for the Dirac distribution on the punctured plane plane, for which a proof  is given in Appendix \ref{AwIvap}, Eq.\; \eqref{dirdir4}, 
\begin{equation*}
\delta\left(\frac{ \bq  }{ \bx}- \bx\right)=\frac{1}{2}\left[\delta\left( \bx-\sqrt{ \bq  }\right) +\delta\left( \bx+\sqrt{ \bq  }\right)\right]\, ,  
\end{equation*}
and where the second Dirac in the r.h.s. has to be taken into account solely  if $\bq=(q,\theta)$ lies in its second Riemann sheet, i.e., for  $2\pi\leq \theta< 4\pi$.
Then, 
\begin{align}
 \label{trUI}  &\mathrm{Tr}\left(U( \bq , \bp )\mathcal{I}\right) =
  \frac{(2\pi)^2 }{{2\,\Vert C_{DM}\psi\Vert}^2} \int_{\R_{\ast}^2} \frac{\ud^2 \bq^{\prime}}{q^{{\prime}2}}
\overline{\psi\left(\frac{\sqrt{ \bq }}{ \bq^{\prime}}\right)} \,\psi\left(\frac{\sqrt{ \bq }}{ \bq^{\prime}}\right)\,e^{\ii \bp \cdot\sqrt{ \bq }} 
   =  \frac{1}{2}e^{\ii \bp \cdot\sqrt{ \bq }}\,,
\end{align}
from the definition of $C_{DM}$ and  after the  last change $ \bq^{\prime}\mapsto \bq^{\prime\prime}= \dfrac{\sqrt{ \bq }}{ \bq^{\prime}}$, which yields $\dfrac{\ud^2 \bq^{\prime}}{q^{{\prime}2}}=\dfrac{\ud^2\bq^{\prime\prime}}{q^{{\prime\prime}2}}$. The equality \eqref{affwigwgt} follows.
 \eprf\\

The unitary operator $\mathcal{I}$ is  nilpotent. From \eqref{trUI} with $ \bq = 1$ and $\bp = \mathbf{0}$,  one checks again that its trace is 
\begin{equation}
\label{traceinv}
\mathrm{Tr}( \mathcal{I})= \frac{1}{2}\,.
\end{equation}
 Note that the two inverse Duflo-Moore operators appearing in the integral \eqref{affbop} simplify to the factor $1/(2\pi)^2$ since $1/\sqrt{Q}\, \mathcal{I} \,1/\sqrt{Q} = \mathcal{I}$.  Section \ref{inversion} is devoted to the study of this important particular case.

\section{Covariant affine integral quantization from weight function}
\label{genform}
\subsection{ General results}
\label{genqres}
We now  establish general formulas for the integral quantization issued from 
a function $\vap( \bq , \bp )$ on $\Gamma \simeq$ SIM$(2)$ yielding the bounded self-adjoint operator $\sfMv$ defined in \eqref{affbop}. 
This allows us to build a family of operators obtained from $\sfMv$ through the $2$-D affine UIR transport:
\begin{equation}
\sfMv( \bq , \bp )=U( \bq , \bp )\sfMv U^{\dag}( \bq , \bp )\,. 
\end{equation}
Then, the corresponding integral quantization \eqref{quantgr} with $G=\mbox{SIM}(2)$ and $\sfM= \sfMv$ is given by:
\begin{equation}
\label{genqvap}
f\mapsto A^{\vap}_f= \int_{\Gamma}\frac{\ud^2  \bq \,\ud^2  \bp }{c_{\sfMv}} \, f( \bq , \bp )\, \sfMv( \bq , \bp )\,,
\end{equation}
where $c_{\sfMv}$ has to be determined from the condition:
\begin{equation}
A^{\vap}_1= \sI\, . 
\end{equation}
As indicated in \eqref{quantform}, the function or  distribution $f$ on $\Gamma$ should allow to define the operator-valued integral  $A^{\vap}_f$  as the sesquilinear form, 
\begin{equation}
\label{affquantform}
B_f(\phi_1,\phi_2):= \int_{\Gamma}\frac{\ud^2  \bq \,\ud^2  \bp }{c_{\sfMv}} \, f( \bq , \bp )\, \left\lg \phi_1\left | \sfMv( \bq , \bp ) \right | \phi_2\right\rg,
\end{equation}
for all $\phi_1$, $\phi_2$ in the dense linear subspace $C_0^{\infty}({\R_{*}^2})$. 

When it is defined in the above sense, the quantization map \eqref{genqvap} is covariant with respect to the unitary
$2$-D affine action $U$:
\begin{equation}
\label{covaff} U( \bq _0, \bp _0) A^\vap_f U^{\dag}( \bq _0, \bp _0) =
A^\vap_{\mathfrak{U}( \bq _0, \bp _0)f}\, ,
\end{equation}
with
\begin{equation}
\label{covaff2}
 \left(\mathfrak{U}( \bq _0, \bp _0)f\right)( \bq , \bp )=
f\left(( \bq _0, \bp _0)^{-1}( \bq , \bp )\right)= f\left(\frac{ \bq }{ \bq _0},\bq_0^{\ast}( \bp 
- \bp _0) \right)\, ,
\end{equation}
$\mathfrak{U}$ being the left regular representation of the affine
group when $f\in L^2(\Gamma, \ud^2  \bq \, \ud^2  \bp )$.
Let us now describe the $A^\vap_f$ as an integral operator in $\calH$ if the function or distribution $f$ allows such a characterization. 
\beprop
Besides the assumptions \ref{assumv} on the function $\vap( \bq , \bp )$, let us suppose that the function $\Omega_{\beta}( \bu)$ defined in \eqref{Omu} obeys
\begin{equation}
\label{Omresun}
0< \Omega_0(1) < \infty\,, 
\end{equation}
and that the function $f$ defines  $ A^{\vap}_f$ as a quadratic form  in the sense of \eqref{affquantform}. Then the action of $ A^{\vap}_f$ on $\phi$ in $C_0^{\infty}(\R_{*}^{2})$ is given in the form of the linear integral  operator
\begin{equation}
\label{acMom3}
(A^{\vap}_f\phi)( \bx  ) = \int_{\R_{\ast}^2}\mathcal{A}^{\vap}_f( \bx  , \bx^{\prime})\,\phi( \bx^{\prime})\,\ud^2   \bx^{\prime}\, , 
\end{equation}
where the kernel $\mathcal{A}^{\vap}_f$ reads as
\begin{align}
\label{kerMomA}
\mathcal{A}^{\vap}_f( \bx  , \bx^{\prime}) &= \frac{2\pi}{c_{\sfMv}}\int_{\mathbb{R}_{\ast}^2}\frac{\ud^2  \bq }{q^2}\,\cMv\left(\frac{ \bx  }{ \bq },\frac{ \bx^{\prime}}{ \bq }\right)\, \hat{f}_p( \bq , \bx^{\prime}- \bx  )\\
\label{kerMomB}  &= \frac{1}{c_{\sfMv}}\, \frac{x^2}{{x^{\prime}}^2}\int_{\mathbb{R}_{\ast}^2}\frac{\ud^2  \bq }{q^2}\,\widehat{\vap}_p\left(\frac{ \bx  }{\vec {x^{\prime}}},- \bq \right)\, \hat{f}_p\left(\frac{ \bx  }{ \bq }, \bx^{\prime}- \bx  \right)\, .
\end{align}
Here $\hat{f}_p$ is the partial Fourier transform of $f$ with respect to the variable $ \bp $ as it was defined in Eq.\;\eqref{parcoure}.
\enprop
\bprf Let $\phi_1$, $\phi_2$ be two elements in the dense linear subspace $C_0^{\infty}(\R_{*}^{2})$. Supposing that the expression $\lg \phi_1 | A^{\vap}_f\ |\phi_2\rg$ makes sense as a quadratic form, we have from  the action  of $U^{\dag}( \bq , \bp )$ on the right and  of $U( \bq , \bp )$ on the left,
\begin{align*}
&\lg \phi_1 | A^{\vap}_f|\phi_2\rg\\\nonumber
&=\int_{\mathbb{R}_{\ast}^2}\ud^2  \bx  \;\overline{\phi_1( \bx  )}\int_{\Gamma}\frac{\ud^2 \bq \,\ud^2 \bp }{c_{\sfMv}}\,f( \bq , \bp )\,\left(U( \bq , \bp )\mathsf{M}^{\vap}U^{\dag}( \bq , \bp )\phi_2\right)( \bx  ) \\\nonumber
&=\int_{\mathbb{R}_{\ast}^2}\ud^2  \bx  \;\overline{\phi_1( \bx  )}\;\int_{\R_{\ast}^2}\ud^2  \bx  ^{\prime}\int_{\Gamma}\frac{\ud^2 \bq \;\ud^2 \bp }{c_{\sfMv}}f( \bq , \bp )\;e^{\ii  \bp . \bx  }\;\ {\mathcal{M}^{\vap}}\left(\frac{ \bx  }{ \bq }, \bx^{\prime}\right)\;e^{-\ii { \bq  ^{\,*} \bp . \bx^{\prime}}}\phi_2( \bq \; \bx^{\prime})\\\nonumber
&=\frac{1}{2\pi}\int_{\mathbb{R}_{\ast}^2}\ud^2  \bx  \;\overline{\phi_1( \bx  )}\int_{\R_{\ast}^2} \ud^2 \bx^{\prime}\;\int_{\Gamma}\frac{\ud^2 \bq \;\ud^2 \bp }{c_{\sfMv}}f( \bq , \bp )\;e^{-\ii  \bp .( \bq   \bx^{\prime}- \bx  )}\;\frac{x^2}{q^2{x^{\prime}}^2}\;\widehat{\vap}_p\left(\frac{ \bx  }{ \bq  \bx^{\prime}},-\frac{ \bx  }{ \bq }\right)\;\phi_2( \bq \; \bx^{\prime})\\\nonumber
&=\int_{\mathbb{R}_{\ast}^2}\ud^2  \bx  \;x^2\;\overline{\phi_1( \bx  )}\;\int_{\R_{\ast}^2} \frac{\ud^2 \bx^{\prime}}{{x^{\prime}}^2}\;\int_{\R_{\ast}^2}\frac{\ud^2 \bq }{c_{\sfMv}}\frac{1}{q^2}\hat{f_{p}}( \bq , \bq   \bx^{\prime}- \bx  )\;\widehat{\vap}_p\left(\frac{ \bx  }{ \bq  \bx  },-\frac{ \bx  }{ \bq }\right)\phi_2( \bq \; \bx^{\prime})\, , 
\end{align*} 
After the  change of variable $ \bx^{\prime}\rightarrow \bx^{\prime\prime}= \bq  \bx^{\prime}$ 
we get,
\begin{align*}
&\lg \phi_1 | A^{\vap}_f|\phi_2\rg\\\nonumber
&=\int_{\mathbb{R}_{\ast}^2}\ud^2  \bx  \;\overline{\phi_1( \bx  )}\;x^2\int_{\R_{\ast}^2} \frac{\ud^2\bx^{\prime\prime}}{{x^{\prime\prime}}^2}\phi_2(\bx^{\prime\prime})\,\frac{1}{c_{\sfMv}}\int_{\R_{\ast}^2}\frac{\ud^2 \bq \;}{q^2}\hat{f_p}( \bq ,\bx^{\prime\prime}- \bx  )\;\widehat{\vap}_p\left(\frac{ \bx  }{\bx^{\prime\prime}},-\frac{ \bx  }{ \bq }\right)\\ \nonumber
&\equiv\int_{\mathbb{R}_{\ast}^2}\ud^2  \bx  \;\overline{\phi_1( \bx  )}\,\int_{\mathbb{R}_{\ast}^2}\ud^2  \bx^{\prime}{\mathcal{A}^{\vap}_{f}}\left( \bx  , \bx^{\prime}\right)\,\phi_2( \bx^{\prime}).
\end{align*}
Therefore
\begin{align*}
\mathcal{A}^{\vap}_f( \bx  , \bx^{\prime}) = \frac{1}{c_{\sfMv}}\frac{x^2}{{x^{\prime}}^2}\int_{\R_{\ast}^2}\frac{\ud^2 \bq \;}{q^2}\hat{f_p}( \bq , \bx^{\prime}- \bx  )\;\widehat{\vap}_p\left(\frac{ \bx  }{ \bx^{\prime}},-\frac{ \bx  }{ \bq }\right)\, ,
\end{align*}
and proceeding with $ \bq  \rightarrow  \bq^{\prime}=\dfrac{ \bx  }{ \bq },\; \dfrac{\ud^2 \bq }{q^2}=\dfrac{\ud  \bq^{\prime}}{{q^{\prime}}^2}$, we have the alternative formula,
\begin{align*}
\mathcal{A}^{\vap}_f( \bx  , \bx^{\prime}) = \frac{1}{c_{\sfMv}}\, \frac{x^2}{{x^{\prime}}^2}\int_{\mathbb{R}_{\ast}^2}\frac{\ud^2  \bq }{q^2}\,\widehat{\vap}_p\left(\frac{ \bx  }{\bx^{\prime}},- \bq \right)\, \hat{f}_p\left(\frac{ \bx }{\bq}, \bx^{\prime}- \bx  \right).\,
\end{align*}
The constant $c_{\sfMv}$ can be fixed by choosing $f( \bq , \bp )=1$, so $\hat{f}_p\left(\frac{ \bx  }{ \bq }, \bx^{\prime}- \bx  \right)=2\pi\delta( \bx^{\prime}- \bx  )$, and then,
\begin{align*}
\mathcal{A}^{\vap}_1( \bx  , \bx^{\prime}) = \frac{2\pi}{c_{\sfMv}}\,\int_{\mathbb{R}_{\ast}^2}\frac{\ud^2  \bq }{q^2}\,\widehat{\vap}_p\left(\frac{ \bx  }{\bx^{\prime}},- \bq \right)\delta( \bx^{\prime}- \bx  )\equiv \delta( \bx^{\prime}- \bx  )\, ,
\end{align*}
which gives,
\begin{equation}
\label{cnresun}
c_{\sfMv} =  2\pi \int_{\mathbb{R}_{\ast}^2}\frac{\ud^2  \bq }{q^2}\,\widehat{\vap}_p\left(1,- \bq \right) \equiv  2\pi\, \Omega_0(1)< \infty\, , 
\end{equation} 
which justifies \eqref{Omresun}.
\eprf

\subsection{Quantization examples}

\subsubsection{Position dependent functions $f$}
Suppose that $f$ does not depend on $ \bp $, $f( \bq , \bp )\equiv u( \bq )$. From 
\begin{equation*}
\hat{f}_p( \bq , \bx^{\prime}- \bx  )= 2\pi\,u( \bq )\,\delta( \bx^{\prime}- \bx  )\, , 
\end{equation*}
and after integration one obtains for \eqref{kerMomA}
\begin{align}
\label{Auq}
\nonumber
\mathcal{A}^{\vap}_{u( \bq )}( \bx  , \bx^{\prime})&= \frac{4\pi^2}{c_{\sfMv}}\,\delta( \bx  - \bx^{\prime})\int_{\R_{\ast}^2}\frac{\ud^2 \by}{y^2}\,\cMv\left(\frac{ \bx  }{\by},\frac{ \bx  }{\by}\right)\,u(\by)\\
 &= \frac{4\pi^2}{c_{\sfMv}}\,\delta( \bx  - \bx^{\prime})\int_{\R_{\ast}^2}\frac{\ud^2 \by}{y^2}\,\cMv\left(\by,\by\right)\,u\left(\frac{ \bx  }{\by}\right)\,.
\end{align}
 Thus,  the quantum version of the function $u( \bq )$  is the multiplication operator
 \begin{equation}
\label{quq}
A^{\vap}_{u( \bq )}=  \frac{4\pi^2}{c_{\sfMv}}\int_{\R_{\ast}^2}\frac{\ud^2 \by}{ y^2}\,\cMv\left(\by,\by\right)\,u\left(\frac{ \bQ }{\by}\right)
=  \frac{2\pi}{c_{\sfMv}}\int_{\R_{\ast}^2}\frac{\ud^2 \by}{y^2}\,\widehat{\vap}_p(1,-\by)\,u\left(\frac{ \bQ }{\by}\right)\, . 
\end{equation}
This can be viewed as a convolution,
on the multiplicative group $\R^{2}_{*}=\R^{+}_{*}\times \mbox{SO}(2)$, of the function $u( \bx  )$ with $ \frac{2\pi}{c_{\sfMv}}\,\widehat{\vap}_p(1,- \bx  )$ followed by the replacement $ \bx  \mapsto  \bQ $. This \textit{$2$D-affine convolution product} is defined by
\begin{equation}
\label{conv2d}
(f_1\ast_{aff} f_2)( \bx  )=\int_{\R_{\ast}^2}\frac{\ud^2 \bx^{\prime}}{{x^{\prime}}^2}\,f_1( \bx^{\prime})f_2\left({\frac{ \bx  }{ \bx^{\prime}}}\right)\,.
\end{equation}
With this notation, we write the result as
\begin{equation}
A_{u( \bq )}\phi( \bx  )=\frac{1}{c_{\sfMv}}(w*_{aff}u)( \bx  ) \phi( \bx  )\, , 
\end{equation}
where $w( \bx  )=\widehat{\vap}_p(1,- \bx  )$.

Some interesting particular cases occur here: 
\begin{enumerate}
\item $u( \bq )$ is a simple power of $q$, say $u( \bq )= q^{\beta}$. Then we have \begin{equation}
\label{Aqbeta}
A^{\vap}_{q^{\beta}}=  \frac{2\pi}{c_{\sfMv}}\int_{\R_{\ast}^2}\frac{\ud^2 \by}{y^{2+\beta}}\,\widehat{\vap}_p(1,-\by)\,Q^{\beta}=  \frac{\Omega_{\beta}(1)}{\Omega_0(1)}\, Q^{\beta}\, , 
\end{equation}
together with necessary conditions of convergence for the integral $\Omega_{\beta}(1)$.
\item  $u( \bq )= \bq $ stands for the  classical vector position in the plane, $\Omega_{(2,0,1)}(1)=0,$
\begin{align}
\label{Aq1-2}
A^{\vap}_{ \bq }
&=  \frac{2\pi}{c_{\sfMv}}\int_{\R_{\ast}^2}\frac{\ud^2 \by}{y^{2}}
\,\widehat{\vap}_p(1,-\by)\,\frac{ \bQ }{\by}\\\nonumber
&
=  \frac{2\pi}{c_{\sfMv}}\int_{\R_{\ast}^2}\frac{\ud^2 \by}{y^{4}}
\,\widehat{\vap}_p(1,-\by)\, (\by\cdot \bQ ,\by\times \bQ )\\\nonumber
&=  \frac{2\pi}{c_{\sfMv}} (\Omega_{(2,1,0)}(1) Q_1+\Omega_{(2,0,1)}(1) Q_2, \Omega_{(2,1,0)}(1) Q_2-\Omega_{(2,0,1)}(1)Q_1)\\\nonumber
&=  \frac{2\pi}{c_{\sfMv}}{\begin{pmatrix}
 \Omega_{(2,1,0)}(1) & \Omega_{(2,0,1)}(1) \\       
 -\Omega_{(2,0,1)}(1) & \Omega_{(2,1,0)}(1) \\      
\end{pmatrix}}
{\begin{pmatrix}
Q_1 \\       
Q_2 \\      
\end{pmatrix}}\, , 
\end{align} 
where we have generalized the notation \eqref{Omu} for the following integrals
\begin{equation}
\label{Omuqq}
\Omega_{\beta, \nu_1,\nu_2}( \bu):= \int_{\mathbb{R}_{\ast}^2}\frac{\ud^2 \by}{y^{\beta +2}}\,\widehat{\vap}_p\left( \bu,-\by\right)\,y_1^{\nu_1}\,y_2^{\nu_2}\,, \quad \Omega_{\beta,0,0}( \bu)\equiv \Omega_{\beta}( \bu)\,,
\end{equation}
with $\by= \begin{pmatrix}
      y_1    \\
      y_2  
\end{pmatrix}$.
\end{enumerate}

\subsubsection{Separable functions $f$}
Suppose that $f$ is separable, i.e.,  $f( \bq , \bp )\equiv u( \bq )\,v( \bp )$. The formula \eqref{kerMomB} simplifies to 
\begin{align}
\label{PosMomA}
&\mathcal{A}^{\vap}_{u( \bq )v( \bp )}( \bx  , \bx^{\prime})\\\nonumber
&=  \frac{1}{c_{\sfMv}}\, \frac{x^2}{{x^{\prime}}^2}\int_{\R_{\ast}^2}\frac{\ud^2  \bq }{q^2}\,\hat v( \bx^{\prime}- \bx  )\,\widehat{\vap}_p\left(\frac{ \bx  }{ \bx^{\prime}},- \bq \right)\,u\left(\frac{ \bx  }{ \bq }\right)\,.\\\nonumber
&=  \frac{1}{c_{\sfMv}}\,\hat v( \bx^{\prime}- \bx  )\, \frac{x^2}{{x^{\prime}}^2}\int_{\R_{\ast}^2}\frac{\ud^2  \bq }{q^2}\,\widehat{\vap}_p\left(\frac{ \bx  }{ \bx^{\prime}},- \bq \right)\,u\left(\frac{ \bx  }{ \bq }\right)\,.\\\nonumber
&=  \frac{1}{c_{\sfMv}}\,\hat v( \bx^{\prime}- \bx  )\, \frac{x^2}{{x^{\prime}}^2}\, \left(\widehat{\vap}_p\left(\frac{ \bx  }{ \bx^{\prime}},-.\right)\,\ast_{aff} u\right)( \bx )\,, \end{align}
where the dot inside parenthesis stands for the integration variable in the affine convolution product, 
and $\hat v$ is the Fourier transform of $v$.

In particular, if $v( \bp )=p_i^{n_i}$, i=1 or 2,
\begin{equation}
\hat{v}(\by)={\frac{1}{2\pi}}\int_{\R_{\ast}^{2}}\ud^2 \bx   \,e^{-\ii \by\cdot  \bx  }x_i^n
=2\pi\;{(-\ii)}^n\frac{\partial^{n}}{\partial y_i^n}\delta(\by)
=2\pi\;{(\ii)}^n\delta(y_{3-i})\frac{\partial^{n}}{\partial y_i^n}\delta(y_i)\, .
\end{equation}
Hence, 
\begin{align}
{\mathcal{A}^{\vap}}_{u( \bq )\, p_i^n}&=\frac{{\ii}^n\;2\pi}{c_{c_{\sfMv}}}\frac{x^2}{{x^{\prime}}^2}
\delta({{x}^{\prime}}_{3-i}-x_{3-i})\,\delta^{(n_i)}_{x^{\prime}_i}({x^{\prime}}_i-x_i)
\, \frac{x^2}{{x^{\prime}}^2}\, \left(\widehat{\vap}_p\left(\frac{ \bx  }{ \bx^{\prime}},-.\right)\,\ast_{aff}u\right)(\bx)\, . 
\end{align}
Applying the corresponding operator on $\phi \in C^{\infty}\left(\R^2_\ast\right)$ yields the expansion in powers of the operator $P_i = -\ii\partial/\partial x_i$:

 \begin{equation}
 \label{Aupin}
 A_{u( \bq )p_{i}^n}\phi( \bx  )= \frac{2\pi Q^2}{c_{\sfMv}}\sum_{s=0}^{n}\binom{s}{n}(-\ii)^{n-s}\frac{\partial^{n-s}}{\partial {{x^{\prime}}_i}^{n-s}}
 \left[\frac{1}{{x^{\prime}}^2}\left(\widehat{\vap}_p\left(\frac{ \bQ }{ \bx^{\prime}},-.\right)\,\ast_{aff}u\right)(\bQ)\right]_{ \bx^{\prime}= \bQ }\,P_i^{s}\phi( \bx  )\,. 
 \end{equation}
For the lowest powers $s$, this formula particularizes as follows. 
\begin{enumerate}
\item  $f( \bq , \bp )=u( \bq )  \bp $
\begin{equation}
\begin{split}
A_{u( \bq ) \bp } &= \frac{2\pi}{c_{\sfMv}}(\widehat{\vap}_p(1,-\cdot)\ast_{aff}u)( \bQ )\, \bP   +\\
&+ \frac{4\pi \ii}{c_{\sfMv}}\frac{1}{\bQ^{\ast}} (\widehat{\vap}_p(1,-\cdot)\ast_{aff}u)( \bQ )
+ \frac{2\pi \ii}{c_{\sfMv}}\frac{1}{\bQ^{\ast}} ((\pmb{\nabla} \widehat{\vap}_p)(1,-\cdot)\ast_{aff}u)( \bQ)\, .
\end{split} 
\end{equation}
\item  For $u( \bQ )=1$, that is, $f( \bq , \bp )= \bp $, then the previous one simplifies to
\begin{equation}
A_{ \bp } =  \bP  + \frac{\ii}{\bQ^{\ast}}\left(2+\frac{\pmb{\Omega}^{(1)}_0(1)}{\Omega_0(1)}\right)\, , 
\end{equation}
where we have introduced another family of integrals:
\begin{equation}
\pmb{\Omega}^{(1)}_{\beta}( \bx  )=\int_{\R_{\ast}^2}\frac{d^2 \bq  }{q^{2+\beta}}(\pmb{\nabla} \widehat{\vap}_p)( \bx  ,- \bq )\,.
\end{equation}
\item   For $u( \bQ )=1$ and $f( \bq , \bp )= \bp  ^{\,2}$ (kinetic energy), \eqref{Aupin} yields:
 \begin{equation}
 \label{Ap2}
 A_{ \bp  ^{\,2}}= \bP  ^2+2\ii\left[2+\frac{\mathbf{e}_1\cdot\pmb{\Omega}^{(1)}_{0}(1)}{\Omega_{0}(1)}\right]\frac{ \bQ }{Q^2}\cdot  \bP   -\left[4+4\frac{\mathbf{e}_1\cdot\pmb{\Omega}^{(1)}_{0}(1)}{\Omega_{0}(1)}+\frac{\Omega^{(2)}_{0}(1)}{\Omega_{0}(1)}\right]\frac{1}{Q^2}\,,
 \end{equation}
 where the new family of integrals appears:
\begin{equation}
\pmb{\Omega}^{(2)}_{\beta}( \bx  )=\int_{\R_{\ast}^2}\frac{d^2 \bq  }{q^{2+\beta}}(\pmb{\nabla}\cdot\pmb{\nabla} \widehat{\vap}_p)( \bx  ,- \bq )\,. 
\end{equation}
 
 \item $f( \bq , \bp )= \bq \cdot  \bp = q_{1} p_{1}+q_{2} p_{2}$ (generator of dilations in the plane).
\\  By introducing the family of integrals 
\begin{align}
\label{Omuqq1}
&\pmb{\Omega}^{(1)}_{(\beta,\nu_1,\nu_2)}( \bx  )=\int_{\R_{\ast}^2}\frac{d^2 \bq  }{q^{2+\beta}}\,(\pmb{\nabla} \widehat{\vap}_p)( \bx  ,- \bq ){q_1}^{\nu_1}\,q_2^{\nu_2},\\\nonumber
\end{align}
and using\eqref{Omuqq}, we get:
\begin{align}
A_{ \bq \cdot  \bp }=\frac{2\pi}{c_{\sfMv}}\left({\Omega}_{(2,1,0)}(1)( \bQ \cdot \bP   +2\ii)+\Omega_{(2,0,1)}  \bQ \times \bP  +\ii(\pmb{\Omega}^{(1)}_{(2,1,0)}(1)\cdot\mathbf{e}_1-\pmb{\Omega}^{(1)}_{(2,0,1)}(1)\cdot\mathbf{e}_2) \right).
\end{align}

\item  $f( \bq , \bp )= \bq \times \bp = q_1p_2-q_2p_1$ (angular momentum).
\\ The result is given  in terms of the integrals \eqref{Omuqq} and \eqref{Omuqq1} as
\begin{equation}
\label{}
A_{ \bq \times  \bp }= 
\frac{2\pi}{c_{\sfMv}}\left(\Omega_{2,1,0}(1) \bQ \times  \bP  +\Omega_{2,0,1}( \bQ \cdot \bP  +2\ii)
- \ii(\pmb{\Omega}^{(1)}_{(2,0,1)}(1)\cdot\mathbf{e}_1+\pmb{\Omega}^{(1)}_{(2,1,0)}(1)\cdot\mathbf{e}_2)\right)\,.
\end{equation}
 \end{enumerate}
 Simplified expressions for the above formulas are always possible  with a suitable choice of the weight function $\vap$. 
 
\section{Quantum phase space portraits}
\label{portrait}
Given a function $\vap( \bq , \bp )$ on the phase space $\Gamma$ and  yielding a self-adjoint unit trace  operator $\sfMv$, the quantum phase space portrait  of an operator $A$ in $\calH$ reads as
\begin{equation}
\label{semclA}
\widecheck{A}( \bq , \bp ):=  \mathrm{Tr}\left(A\,U( \bq , \bp )\,\sfMv\,U^{\dag}( \bq , \bp )\right) = \mathrm{Tr}\left(A\,\sfMv( \bq , \bp )\right)\, . 
\end{equation}
The most interesting aspect of this notion in terms of probabilistic interpretation holds when the operator $A$ is precisely the affine integral quantized version $A_f$ of a classical $f(q,p)$ with the same function $\vap$ (actually we could define the transform with 2 different ones, one for the ``analysis'' and the other for the ``reconstruction''). Then, we get the transform of the type \eqref{lowsymb}
 \begin{equation}
\label{lowsymbv}
\begin{split}
f( \bq , \bp )\mapsto \widecheck f( \bq , \bp ) &\equiv \check{A}_f^{\vap}( \bq , \bp )\\&= \int_{\Gamma}\frac{\ud^2  \bq^{\prime} \, \ud^2  \bp^{\prime}}{c_{\sfMv}}\, f\left( \bq  \bq^{\prime},\frac{ \bp^{\prime}}{ \bq }+ \bp \right)\, \mathrm{Tr}\left(\sfMv( \bq^{\prime}, \bp^{\prime})\sfMv\right)\ . 
\end{split}
\end{equation}
This expression (which can be viewed as a convolution on SIM$(2)$) has the meaning of an averaging of the classical $f$ if the function 
\begin{align}
\label{distvap}
 ( \bq , \bp ) &\mapsto \frac{1}{c_{\sfMv}}\mathrm{Tr}\left(\sfMv( \bq , \bp )\sfMv\right)\, ,
\end{align}
is a true probability distribution on $\Gamma$, i.e. is positive since we know from the resolution of the identity that its integral is 1. 
In fact, this is a kind of Husimi function \cite{husimi40}.
The  expression of $\mathrm{Tr}\left(\sfMv( \bq , \bp )\sfMv\right)$  is easily derived from the kernel expression \eqref{kerMom} and reads as
\begin{equation}
\label{trmommom}
\mathrm{Tr}\left(\sfMv( \bq , \bp )\sfMv\right)= \frac{1}{(2\pi)^2 q^2} \, \int_{\R_{\ast}^2}\ud^2  \bx  \int_{\R_{\ast}^2} \ud^2 \by \, e^{\ii  \bp .( \bx  -\by)}\, \widehat\vap_p\left(\frac{ \bx  }{\by},-\frac{ \bx  }{ \bq }\right)\,\,\widehat\vap_p\left(\frac{\by}{ \bx  },-\by\right)\, . 
\end{equation}
Integrating this expression on $\Gamma$ with the measure $\dfrac{\ud^2  \bq \,\ud^2  \bp }{c_{\sfMv}}$ and using \eqref{Tr2cond} (i.e., $\Omega_{-2}/2\pi = \mathrm{Tr}(\sfMv) = 1$) and \eqref{cnresun}, we get $1$, which means that $\widecheck 1= 1$, as expected. 

For instance, one easily shows that, for the coherent state weight with respect to the fiducial vector $\psi$ (\ref{weightcs}), we have $\mathrm{Tr}\left(\sfMvp( \bq , \bp )\sfMvp\right) = {\vert \langle U( \bq , \bp )\psi|\psi \rangle \vert}^2
$.

The lower symbol of a function $f( \bq , \bp )\equiv u( \bq )$ is  a  function  of $ \bq $ only. 
\begin{equation}
\label{lowuq}
\widecheck u( \bq ) = \int_{\R_{\ast}^2}\frac{\ud^2  \bx }{c_{\sfMv}}\, \widehat\vap_p(1, \bx  ) \left(\widehat\vap_p(1,\cdot)\ast_{\mathrm{aff}}u\right)
( \bq  \, \bx  )\,.
\end{equation}
It is expected more regular.
Note the two marginal integrations allowing the respective distributions on $\R^2_{\ast}$ and $\R^2$:
\begin{align}
\label{distq}
    \bq  \mapsto  \int_{\R^2}\frac{\ud^2  \bp }{c_{\sfMv}}\,  \mathrm{Tr}\left(\sfMv( \bq , \bp )\sfMv\right) &= \frac{1}{c_{\sfMv}}\int_{\R_{\ast}^2}\ud^2 \bx  \, \widehat\vap_p(1, \bx  )\, \widehat\vap_p(1,  \bq \, \bx  )\\
\label{distq1}  &=  \frac{1}{c_{\sfMv}}\int_{\R^2}\ud^2 \mathbf{ k}\, \vap(1,\mathbf{ k})\, \vap(1, - \bq ^{\ast}\,\mathbf{ k})
   \, ,
\end{align}
and, 
\begin{equation}
 \label{distp}  
 \begin{split}
 \bp  \mapsto  &\int_{\R_{\ast}^2}\frac{\ud^2  \bq }{c_{\sfMv}}\,  \mathrm{Tr}\left(\sfMv( \bq , \bp )\sfMv\right)\\ &= - \frac{1}{2\pi c_{\sfMv}} \int_{\R_{\ast}^2}\ud^2  \bx\, \Omega( \bx  )\,\Delta_{ \bu}\left[\vap\left(\frac{1}{ \bx}, \bu\right)\right]_{ \bu= \left(1- \bx^{\ast}\right)\, \bp }\,, 
\end{split}  
 \end{equation}
where $\Omega( \bx  )$ is defined in \eqref{Omu}.

\section{Quantization  and semi-classical portrait through the inversion map } 
\label{inversion}

In this section we investigate the integral quantization and the subsequent semi-classical portrait yielded by the operator $\sfMva=2\mathcal{I}$ introduced in \eqref{MpaW},   where $\mathcal{I}$ is the unitary  inversion operator  defined on $L^2(\R^2,\ud^2  \bx  )$ as it was  introduced in Eq.\;\eqref{invop}. 

\subsection{Quantization through the inversion map}

\beprop
\label{propqinv}
The integral kernel \eqref{kerMomB} of the operator $ A^{\vapa}_f$ which is the quantization of the  function $f( \bq , \bp )$ through the  function $\vapa$ given in \eqref{affwigwgt} has the following expression,
\begin{equation}
\label{intkerqI}
\mathcal{A}_f^{\vapa}( \bx  , \bx^{\prime}) =\frac{1}{2\pi} \widehat f_p\left(\sqrt{ \bx    \bx^{\prime}},  \bx^{\prime}- \bx  \right)\,. 
\end{equation}
\enprop
\bprf
The computation of the partial Fourier transform of the function $\vapa( \bq , \bp )=\dfrac{e^{-\ii\bp \cdot\sqrt{\bq}}}{q}$   yields 
\begin{equation*}
\widehat{\vap}_{a\mathcal{W}; \bp  }\left(\frac{ \bx}{ \bx^{\prime}},- \bq  \right)= \frac{2\pi}{q} \delta\left(  \bq  - \sqrt{\frac{ \bx}{ \bx^{\prime}}}\right)\, .
\end{equation*}
It follows for the integral $\Om( \bu)$  defined by \eqref{Omu}  and   the constant \eqref{cnresun}  the simple values
\begin{equation}
\label{ cMvapa}
\Om( \bu)= \frac{2\pi}{ u}\, , \quad c_{\mathsf{M}^{\vap_{a\mathcal{W}}}}= (2\pi)^2\,, 
\end{equation} 
and so for the kernel \eqref{kerMomB},
\begin{equation*}
\label{kerMomA1}
\mathcal{A}^{\vapa}_f( \bx, \bx^{\prime}) = \frac{1}{2\pi}  \widehat f_p\left(\sqrt{ \bx    \bx^{\prime}},  \bx^{\prime}- \bx  \right)\,.
\,\end{equation*}

\eprf\\
We derive from Proposition \ref{propqinv} the following interesting results holding for this particular type of integral quantization.
\beprop
\label{propqinvpcuv}
\begin{itemize}
  \item[(i)] The quantization of a function of $ \bq $, $f( \bq , \bp )= u( \bq )$ provided by the weight $\vapa$ is   $u( \bQ )$.
 \item[(ii)] Similarly, the quantization of a function of $ \bp $, $f( \bq , \bp )= v( \bp )$, provided by the weight $\vapa$ is   $v( \bP  )$ (in the general sense of pseudo-differential operators yielded by the Fourier transform of tempered distributions),
 \begin{equation}
\label{ AWvp}
\left(A^{\vapa}_{v( \bp )}\,\psi\right)( \bx  )=  \frac{1}{2\pi}  \int_{\R^2}\,\ud^{2} \bp  \, e^{\ii  \bp \cdot  \bx  }\, v( \bp )\,\widehat \phi( \bp )\, .
\end{equation}
  \item[(iii)] More generally, the quantization of a separable function $f( \bq , \bp )= u( \bq )\,v( \bp )$ provided by the weight $\vapa$ is the integral operator
  \begin{equation}
\label{Auvinvq}
\left(A^{\vapa}_{u( \bq )v( \bp )}\,\psi\right)( \bx  )= \frac{1}{4\pi}  \int_{\R_{\ast}^2}\,\ud^{2}\, \bx^{\prime} \,\widehat v( \bx^{\prime}- \bx  )\, u\left(\sqrt{ \bx  \, \bx^{\prime}}\right)\, \psi( \bx^{\prime})\, .
\end{equation}
 \item[(iv)] In particular, the quantization of $u( \bq )\,{p}_l^n$, $l=1,2,$, $n\in \N$, yields the symmetric operator,
 \begin{equation}
\label{Apnuinvq}
A^{\vapa}_{u( \bq )p_l^n}= \sum_{k=0}^n \binom{n}{k} \,(-\ii)^{n-k}\partial^{(n-k)}_{x^{\prime}_l} \,\left[u\left(\sqrt{ \bx  \, \bx^{\prime}}\right)\right]_{ \bx  = \bx^{\prime}= \bQ }\, P_l^k\, , 
\end{equation}
and for the dilation,
\begin{equation}
\label{Apqinvq}
A^{\vapa}_{ \bq \cdot  \bp }= D\, .
\end{equation}
\end{itemize}
\enprop
\bprf The proof is made through a direct application of Eq.\;\eqref{intkerqI} and application of elementary distribution theory.
 \eprf\\
Therefore, this affine integral  quantization is the exact counterpart of the Weyl-Wigner  integral quantization \cite{bergaz14} and can be termed as canonical as well.  However, the choice of such a procedure would lead to three difficulties, at least,
\begin{enumerate}
  \item Since the quantization of the kinetic energy of the free particle on the punctured plane is just 
  \begin{equation}
\label{aWp2}
A^{\vapa}_{ \bp  ^{\,2}}=  \bP  ^2\, , 
\end{equation}
and thus does not produce an essentially self-adjoint operator, the corresponding  quantum dynamics depends on boundary conditions at the origin $ \bx  =0$. There exists an irreducible ambiguity since different physics are possible on the quantum level. 
  \item No classical singularity is cured on the quantum level since
  \begin{equation}
\label{aWuqvp}
   A^{\vapa}_{u( \bq )}  = u( \bQ )\, , \qquad A^{\vapa}_{v( \bp )}  = v( \bP  )\, ,
\end{equation}
a feature of the Weyl-Wigner  integral quantization as well.
\item The semi-classical portraits of quantum operators along Eqs.\;\eqref{semclA} and \eqref{lowsymbv} cannot be given a probabilistic interpretation, a feature of the Weyl-Wigner  integral quantization as well. 
\end{enumerate}
The last point is developed below for rank-one operators  $|\phi\rg\lg \phi|$, i.e. pure states. 

\subsection{Affine Wigner-like quasi-probability}
The  affine Wigner-like quasi-probability $\mathcal{AW}_{\phi}$ corresponding to the  state $\phi$ is the application of the general expression \eqref{semclA} to the projector $|\phi\rg\lg \phi|$:
\begin{equation}
\label{AWpsi}
\mathcal{AW}_{\phi}( \bq , \bp )
:=\lg \phi| \sfM^{\vap_{a\mathcal{W}}}( \bq , \bp )|\phi \rg= 2\int_{\R_{\ast}^2}\ud^2  \bx   \,\overline{\phi\left( \bx  \right)} \, e^{\ii  \bp \cdot\left( \bx  -\frac{ \bq  ^{\,2}}{ \bx  }\right)}\frac{q^2}{x^2}\, \phi\left(\frac{ \bq  ^{\,2}}{ \bx  }\right)\,.
\end{equation}
\beprop
Let us consider a unit norm vector  $\phi \in L^2(\R_{\ast}^2, \ud^2  \bx  ) \cap L^1(\R_{\ast}^2, \ud^2  \bx  )$ and the corresponding quasi-probability distribution $\mathcal{AW}_{\phi}( \bq , \bp )$. The latter verifies the following properties.
\begin{itemize}
\item[(i)] It is real
\begin{equation} 
\overline{\mathcal{AW}_{\phi}( \bq , \bp )}=\mathcal{AW}_{\phi}( \bq , \bp )\,.
\end{equation} 
\item[(ii)] It is a quasi-probability,
\begin{equation}
\int_{\Gamma}\frac{\ud^2  \bq \;\ud^2  \bp }{(2\pi)^2}\mathcal{AW}_{\phi}( \bq , \bp )=1\,.
\end{equation}
\item[(iii)] It satisfies the correct marginalization with respect to  variables $q$ and  $p$ respectively,
\begin{align}
\label{marq}
\int_{\R_{\ast}^2}\frac{\ud^2  \bq }{4\pi^2} \mathcal{AW}_{\phi}( \bq , \bp )
=|\hat\phi( \bp )|^2\,, \quad \hat\phi( \bp )= \frac{1}{2\pi}\int_{\R_{\ast}^{2}} \ud^2  \bq \;e^{-\ii  \bp \cdot \bq }\, \phi( \bq )\,,
\end{align}
\begin{align}
\label{marp}
\int_{\R^2}\frac{\ud^2  \bp }{4\pi^2} \mathcal{AW}_{\phi}( \bq , \bp ) 
=\vert\phi( \bq )\vert^2\,.
\end{align}
\end{itemize}
\enprop
\bprf
(i) and (ii) are  direct consequences of the fact that the operator $\sfM^{\vap_{a\mathcal{W}}}( \bq , \bp )$ is self-adjoint and unit trace. (iii) results from elementary integral calculus on \eqref{AWpsi} with changes of variable  and Fourier transforms in the plane. Fubini applies due to the assumption that $\phi \in L^1(\R^2, \ud^2  \bx  )$, and 
$$
\int_{\R_{\ast}^2} \ud^2  \bq  \,\left\vert\mathcal{AW}_{\phi}( \bq , \bp ) \right\vert 
\leq  \Vert \phi\Vert_{L^1}^2\,.
$$
\eprf\\
Note that the marginal \eqref{marp} reflects the existence of the singularity at the origin of the punctured plane. 

\subsection{Semi-classical portrait with affine Wigner weight}
For the affine Wigner weight \eqref{affwigwgt} the formula \eqref{trmommom} is not really manageable. The direct calculation starting from \eqref{trace1} yields:
\begin{equation}
\label{distaffwigner}
   \mathrm{Tr}\left(\sfMva( \bq , \bp )\sfMva\right) = 
\delta(\bq - 1) \, \int_{\R_{\ast}^2}\frac{\ud^2  \bx}{x^2} \, e^{\ii \bp\cdot\left(\bx - \frac{1}{\bx}\right)}\,,
\end{equation} 
where the remaining divergent integral should be understood after regularisation as a distribution

We derive from \eqref{lowsymbv} and \eqref{distaffwigner} the semi-classical portrait of the operator $A^{\vapa}_{f}$ reads as the integral transform:
\begin{equation}
\label{affWiglow}
\widecheck f( \bq , \bp )  = \frac{1}{\pi} \int_{\R_{\ast}^2}\frac{\ud^2 \bx}{x^2}\, e^{\ii\bp\cdot\left(\bx-\frac{1}{\bx}\right)}\,\widehat{f}_p\left(\bq,\bx-\frac{1}{\bx}\right)\,.
\end{equation}
 In particular, the semi-classical portrait or lower symbol of the quantization of a function of $ \bq $ alone, $f( \bq , \bp )= u( \bq )$,  is  this function:
\begin{equation}
\label{spuq}
\widecheck{u}( \bq ) = u( \bq )\,.
\end{equation}

On the other hand,  we get for a function of $ \bp$ alone, $f( \bq , \bp )= v( \bp )$:
\begin{equation}
\label{spvp}
\widecheck{v}( \bp ) = \frac{1}{\pi} \int_{\R_{\ast}^2}\frac{\ud^2 \bx}{x^2}\, e^{\ii\bp\cdot\left(\bx-\frac{1}{\bx}\right)}\,\widehat{v}\left(\bx-\frac{1}{\bx}\right)= v(\bp)\,.
\end{equation}

\section{Conclusion}
\label{conclu}
In this paper we have explored the possibilities offered by the $2$-D affine covariant integral quantization beyond 
the cases of the open half-line \cite{gazmur16} and affine coherent states.  The latter was considered in \cite{gazkoimur17} and is reviewed in Appendix \ref{QaffCSth}.  
Such a generalization enables us to apply the integral quantization of the motion of a point particle in the punctured plane $\R_{*}^2$, a case  for which the standard Weyl-Heisenberg approach might be viewed as not appropriate.  The corresponding position-momentum phase space is the manifold $\R_{*}^2\times\R^2$ whose the group structure makes it isomorphic to SIM$(2)$.  Hence,  the central objects are the affine group SIM$(2)$, the weight function $\vap( \bq , \bp )$ defined on  $\R_{\ast}^2\times\R^2\sim \mathrm{SIM}(2)$ and its partial Fourier transform with respect to the momentum variable $ \bp$. This $2$-D affine quantization is implemented on functions (or distributions) on the phase space $\R_{\ast}^2\times\R^2$.  

In our opinion, the most attractive meaning to be given to the weight function lies in its partial Fourier transform through its appearance in   the semi-classical expression \eqref{trmommom}. In the case of the Weyl-Heisenberg integral quantization \cite{gazeau_intsympro18}, an interpretation was given  in terms of probability distribution  for the difference of two vectors in the phase plane, viewed as independent random variables. It would be interesting to find a similar interpretation in terms of probability distribution for the product $(\bq,\bp)\, (\bq^{\prime},\bp^{\prime})^{-1}$ of two  group elements of SIM$(2)$ also viewed as independent random variables.

In  our approach, semi-classical portraits are defined in a systematic manner and then the quasi-probability distributions can be generalized in a unified procedure 
reproducing the well-known Wigner and Husimi functions. 
Moreover, we can investigate the modification of the quasi-distributions  due to the non-trivial geometry of the phase space.

In a parent paper, we will consider some applications of the above formalism to the dynamics of physical systems for which the origin of the configuration space is forbidden, like we have for the Aharonov-Bohm model sketched in the introduction,  or the rotating frame. Indeed, through a suitable choice of the weight function, our procedure allows to regularise on the quantum level the origin of the punctured plane viewed as a classical singularity, e.g., see the related discussion following Eq.\; \eqref{qkinener}. 

On the mathematical side, we may extend our quantisation procedure  for affine symmetries  to higher dimensions. In $n$ dimensions, the group SIM$(n)$
is the semidirect product
\begin{equation}
\label{SIMn}
\mathrm{SIM}(n) = \left(\R^{+}_{\ast} \times \mathrm{SO}(n)\right)\ltimes \R^n\, .
\end{equation}
The corresponding phase space is the left  coset
\begin{equation}
\label{phspn}
\mathrm{SIM}(n)/ \mathrm{SO}(n-1) \simeq   \left(\R^{+}_{\ast} \times \mathbb{S}^{n-1}\right)\times \R^n \simeq  \R^{n}_{\ast}\times \R^n\, .
\end{equation}
The application of the integral quantization based on a notion of square-integrability of UIR of a group  with respect to a subgroup (see \cite{aagbook13} and references therein) can be adapted to the above situation to produce regularizing quantizations from  weight functions on \eqref{phspn}  akin to the content of the present work. We notice that the only cases for which the manifold \eqref{phspn} is also a group hold for $n=1$, $n=2$, and $n=4$. The latter case reads as
\begin{equation}
\label{phspngr}
\mathrm{SIM}(4)/ \mathrm{SO}(3) \simeq   \left(\R^{+}_{\ast} \times \mathbb{S}^{3}\right)\ltimes \R^4 =  \H_{\ast}\ltimes \H\, .
\end{equation}
where $\H$ is the quaternion field. Composition law and inverse for this group are given by:
\begin{equation}
\label{SIM4}
(\bq,\bp)(\bq^{\prime},\bp^{\prime})= \left(\bq\bq^{\prime}, {\overline\bq}^{\,-1}\bp^{\prime} + \bp \right)\, , \quad (\bq,\bp)^{-1}= \left(\bq^{-1}, - \overline\bq  \,\bp\right)\, , 
\end{equation} 
where $\overline\bq$ is the quaternionic conjugate of $\bq$. 

\appendix

\section{A   $2D$-Dirac formula}
\label{AwIvap}
We start by calculating the following Fourier transform in the distribution sense. On one hand we have
\begin{equation}
\label{four1}
\int_{\R^2} \ud^2  \bp\, e^{\ii \bp\cdot \left(\frac{ \bq  }{ \bx}- \bx\right) }= 4\pi^2 \delta\left(\frac{ \bq  }{ \bx}- \bx\right)\,.
\end{equation}
Alternatively, we first change the argument of the exponential as
\begin{equation}
\label{expchange}
 \bp\cdot \left(\frac{ \bq  }{ \bx}- \bx\right) =  \bp\cdot \frac{ \bq  }{ \bx}- \bp\cdot  \bx=  \frac{ \bp}{ \bx^{\ast}}\cdot  \bq  - \bp\cdot  \bx= \frac{\mathrm{R}(\theta) \bp}{x}\cdot  \bq  - \bp\cdot  \bx\,,
\end{equation}
with $x=(x,\theta ) $.
Now we change 
\begin{equation}
\label{chvar1}
\bp^{\prime}= \frac{\mathrm{R}(\theta) \bp}{x}\, ,\quad \ud^2\bp^{\prime}= \frac{\ud^2 \bp}{x^2}\,,
\end{equation} 
to obtain finally
\begin{equation}
\label{four2}
\int_{\R^2} \ud^2  \bp\, e^{\ii \bp\cdot \left(\frac{ \bq  }{ \bx}- \bx\right) }= x^2\int_{\R^2} \ud^2 \bp^{\prime}\, e^{\ii\bp^{\prime}\cdot \left( \bq  - \bx^2\right) }=4\pi^2x^2\delta\left( \bq  - \bx^2\right)\,. 
\end{equation}
Thus
\begin{equation}
\label{dirdir2}
\delta\left(\frac{ \bq  }{ \bx}- \bx\right)=x^2\delta\left( \bx^2- \bq  \right)\,.
\end{equation}
Now, we must be cautious with square roots of $\bq$ as they appear above as roots of $\bx^2- \bq=0$. We should not forget that $\bx$ and $\bq$ are confined to the punctured plane  $\R_\ast^2 \sim \C_\ast$. So, let us proceed with the change of variables
\begin{equation}
\label{yx2}
\bx = (x,\theta) \mapsto \by = (y,\omega) = \bx^2= (x^2,2 \theta)\, , \quad \theta \in [0,2\pi) \Rightarrow \omega \in [0,4\pi)\,.
\end{equation}
 Thus we deal here with the two  sheets of the Riemann surface for the variable $\by$:
 \begin{equation}
\label{riemannsh}
\mathcal{R}_1:= \{ \by = (y,\omega)\,,\, 0\leq \omega < 2\pi\}\, , \quad \mathcal{R}_2:= \{ \by = (y,\omega)\,,\, 2\pi\leq \omega < 4\pi\}\, ,
\end{equation}
with the cut $\{\by=-y\,,\, y\in \R^{+}\}$.
From
\begin{equation*}
\label{intxy}
\int_{\R^{2}_{\ast}} \ud^2\bx= \int_0^{+\infty} x\ud x\int_0^{2\pi}\ud\theta= \frac{1}{2}\int_0^{+\infty}\ud y\int_0^{4\pi}\ud\omega \, ,
\end{equation*} 
we derive the decomposition formula
\begin{equation}
\label{decform}
\int_{\R^{2}_{\ast}} \ud^2\bx= \frac{1}{2}\left[\int_{\mathcal{R}_1} \frac{\ud \by}{y} +\int_{\mathcal{R}_2} \frac{\ud \by}{y}\right]\,.
\end{equation}
This formula gives a precise sense to the Dirac distribution $\delta\left( \bx^2- \bq  \right)$:
\begin{equation}
\label{dirdir3}
\int_{\R^{2}_{\ast}} \ud^2\bx\, \delta\left( \bx^2- \bq  \right)\,\varphi(\bx)= \frac{1}{2q}\left[ \varphi(\sqrt{\bq}) + \varphi(-\sqrt{\bq})\right]\, ,
\end{equation} 
for all test function in some suitable space, e.g. $C^{\infty}$ functions on $\R^{2}_{\ast}$ which rapidly decrease at $0$ and $\infty$, and where $\pm\sqrt{\bq}$ are defined for $\bq =(q,\theta)$, $\theta\in [0, 2\pi)$, as
\begin{equation}
\label{defpmsq}
 \sqrt{\bq}:= \left.\bq^{1/2}\right\vert_{\bq\in \mathcal{R}_1}= (\sqrt{q}, \theta/2)\, , \quad -\sqrt{\bq}:= \left.\bq^{1/2}\right\vert_{\bq\in \mathcal{R}_2}\, .
\end{equation}
Eventually, we get from \eqref{dirdir2} a formula relevant to the present paper:
\begin{equation}
\label{dirdir4}
\int_{\R^{2}_{\ast}} \ud^2\bx\, \delta\left( \bx- \frac{\bq}{\bx}  \right)\,\varphi(\bx)= \frac{1}{2}\left[ \varphi(\sqrt{\bq}) + \varphi(-\sqrt{\bq})\right]\, ,
\end{equation}

\section{Quantization with $2$-D  affine Coherent States (ACS) : a summary }
\label{QaffCSth}

In this appendix, which completes the technical content of \cite{gazkoimur17}, we particularize the general method of $2$-D  affine covariant integral quantisation to the case when the quantiser operator is the projector $|\psi\rg\lg \psi|$ on an admissible fiducial vector, defined by the  function \eqref{weightcs}, as it was described in Subsection \ref{covcs}. 

\subsection{Quantization with ACS}
\label{QaffCS}
Let us  implement the  integral quantization scheme described in this paper by
restricting the method  to the specific case of the rank-one density operator or
projector $\sfM^{\vap_{\psi}}=|\psi\rangle\left\langle \psi\right|$ where $\psi$
is a unit-norm admissible state, or fiducial vector, or wavelet, i.e., is in 
$L^{2}({\R_{\ast}^2},\mathrm{d^2} \bx  ) \cap L^{2}\left({\R_{\ast}^2},\dfrac{\mathrm{d^2} \bx  }{x^2}\right)$. We have from \eqref{Omu} and \eqref{acsvap}:
\begin{equation}
\label{Omdbeta}
\Omega( \bu)(\equiv \Omega_{0}( \bu))=\frac{2\pi}{u^2}  \int_{\R_{\ast}^2}\frac{\ud^2  \bq }{q^2}\, \psi( \bq )\, \overline{\psi\left(\frac{ \bq }{ \bu}\right)}\, , \quad \Omega_{\beta}(1)=2\pi\int_{\R_{\ast}^2}\frac{\ud^2  \bq }{q^{2+\beta}}\,\vert \psi( \bq )\vert^2\, . 
\end{equation}
\beprop
With implicit assumptions on the existence of derivatives of $\psi$ and of their (square)  integrability, we have 
\begin{equation}
\label{Ompsi1a} 
(\pmb{\nabla} \Om)(1)= -2\Omega(1)-i2\pi\left\langle \frac{ \bP  }{ \bQ }\psi|\psi \right\rangle\, ,
\end{equation}
\begin{equation}
\label{Ompsi1b} 
(\Delta \Omega)(1))=4\Omega(1)+8\ii\pi \left\langle \frac{ \bP}{ \bQ }\psi|\psi\right\rangle-2\pi\left\langle  \bP  ^{2}\psi|\psi\right\rangle\, ,
\end{equation} 
which implies for $\psi$ real:
\begin{equation}\label{psireal1} 
2 +\frac{\pmb{\Omega}^{(1)}(1)}{\Omega(1)}=\mathbf{0}\,,
\end{equation}
and, 
\begin{equation}\label{psireal2} 
4+4\frac{\Omega_1^{(1)}(1)}{\Omega(1)}+\frac{\Omega^{(2)}(1)}{\Omega(1)}=
-2\pi\langle  \bP  ^{2}\psi|\psi\rangle\,. 
\end{equation}
\enprop
\bprf
For \eqref{Ompsi1a}, 
\begin{align*}
&\pmb{\nabla}_{ \bu}\left[ \frac{2\pi}{u^2}  \int_{\R_{\ast}^2}\frac{\ud^2  \bq }{q^2}\, \psi( \bq )\, \overline{\psi\left(\frac{ \bq }{ \bu}\right)}\right]_{ \bu =1}= \pmb{\nabla}_{ \bu}\left[ \frac{2\pi}{u^2}  \int_{\R_{\ast}^2}\frac{\ud^2  \bq }{q^2}\, \psi( \bq )\, \overline{\psi\left(\frac{ \bq }{ \bu}\right)}\right]_{ \bu =1}\\\nonumber
&= -2\Omega(1)-2\pi\int_{\R^2_{\ast}}\ud^2 \bq \,  \bq^{-1}\,\pmb{\nabla}_{ \bq }\{\overline{\psi( \bq )}\}\psi( \bq ) = -2\Omega(1)-\ii 2\pi\left\langle \frac{ \bP  }{ \bQ }\psi|\psi \right\rangle\, . 
\end{align*}
For \eqref{Ompsi1b}, 
\begin{align}
\label{Ompsi1bdem}
& \pmb{\nabla}\cdot\pmb{\nabla}_{ \bu}\left[ \frac{2\pi}{u^2}  \int_{\R_{\ast}^2}\frac{\ud^2  \bq }{q^2}\, \psi( \bq )\, \overline{\psi\left(\frac{ \bq }{ \bu}\right)}\right]_{ \bu =1}\\\nonumber
&= \left[\left(\frac{4}{u^2}\right)2\pi\int_{\R_{\ast}^2}\frac{\ud^2  \bq }{q^2}\, \psi( \bq )\, \overline{\psi\left(\frac{ \bq }{ \bu}\right)}+2\left(\frac{-2 \bu}{u^4}\right)\cdot 2\pi\int_{\R_{\ast}^2}\frac{\ud^2  \bq }{q^2}\, \psi( \bq )\,  \frac{- \bq  ^{\,*}}{ \bu^{2*}}\pmb{\nabla}_{\frac{ \bq }{ \bu}}\left(\overline{\psi\left(\frac{ \bq }{ \bu}\right)}\right)+\right.\\\nonumber
&+\left.\frac{2\pi}{u^2}  \int_{\R_{\ast}^2}\frac{\ud^2  \bq }{q^2}\, \psi( \bq )\, \frac{q^2}{u^6}\pmb{\nabla}_{\frac{ \bq }{ \bu}}\cdot \pmb{\nabla}_{\frac{ \bq }{ \bu}}\overline{\psi(\frac{ \bq }{ \bu}})\right]_{ \bu = 1}\\\nonumber
&= 4\times2\pi\int_{\R_{\ast}^2}\frac{\ud^2  \bq }{q^2}\, \psi( \bq )\, \overline{\psi\left( \bq \right)}+4\times2\pi\int_{\R_{\ast}^2}\ud^2  \bq \, \psi( \bq )\,  \frac{ \bq  ^{\,*}}{q^2}\pmb{\nabla}_{ \bq }\overline{\psi\left( \bq \right)}+\\
\nonumber &+2\pi  \int_{\R_{\ast}^2}\ud^2  \bq \, \psi( \bq )\, \pmb{\nabla}_{ \bq }\cdot\pmb{\nabla}_{ \bq }\overline{\psi\left( \bq \right)}\\\nonumber
&=4\Omega(1)+8\ii \pi\left\langle \frac{ \bP  }{ \bQ }\psi|\psi\right\rangle-2\pi\left\langle  \bP  ^{2}\psi|\psi\right\rangle\, . 
\end{align}

Note that $\left\langle \frac{ \bP  }{ \bQ }\psi|\psi \right\rangle$ is purely imaginary and cancels for real $\psi$. 
\eprf\\
Therefore, by applying the general formalism, we recover  a set of results already given in previous works, e.g. in  \cite{berdagama14} in the $1$-D affine case, and in \cite{gazkoimur17} in the $2$-D case.  As was pointed out after stating the orthogonality relations \eqref{orthaffine}, the action of the UIR operators $U( \bq , \bp )$ on $\psi$ produces all affine coherent states, i.e. wavelets, defined as $| \bq , \bp \rangle\
=U( \bq , \bp )|\psi\rangle$. 
Given a certain function $f( \bq , \bp )$ on the phase space, the corresponding affine coherent  state (ACS) quantization reads as
\begin{equation}
\label{quantfaff} f\ \mapsto\
A_{f}=\int_{\Gamma}f( \bq , \bp )| \bq , \bp \rangle\langle
 \bq , \bp |\dfrac{\mathrm{d^2} \bq \mathrm{d^2}\, \bp }{(2\pi)^2 c_{0}}\,,
\end{equation}
which arises from the resolution of the identity
\begin{equation}
\label{affresunit} \int_{\Gamma}| \bq , \bp \rangle\langle
 \bq , \bp |\,\dfrac{\mathrm{d^2} \bq \mathrm{d^2} \bp }{(2\pi)^2 c_{0}}=\sI\,,
\end{equation}
where we adopt for convenience the simplified notations of \cite{berdagama14},
\begin{equation}
\label{cgamma1}
c_{\beta}:=\int_{\R_{\ast}^2}|\psi( \bx  )|^{2}\,\frac{\mathrm{d^2} \bx  }{x^{2+\beta}} =  \frac{\Omega_{\beta}(1)}{2\pi}\,.
\end{equation}
\begin{equation}
\label{cgamma2}
c_{\beta\nu_1\nu_2}:=\int_{\R_{\ast}^2}\,\frac{\mathrm{d^2} \bx  }{x^{2+\beta}}|\psi( \bx  )|^{2}{x_1}^{\nu_1}{x_2}^{\nu_2} = \, \frac{\Omega_{(\beta,\nu_1,\nu_2)}(1)}{2\pi}\,, \quad c_{\beta\,0\,0}\equiv c_{\beta}\,  .
\end{equation}
Thus, a necessary condition  to have \eqref{affresunit} true is
that $c_{0} < \infty$, which implies $\psi(\mathbf{0}) = 0$, a well-known
requirement in wavelet analysis.

Choosing $\psi$ real, and using equations \eqref{psireal1} and \eqref{psireal2} in the quantization formulas established in Subsection \ref{genqres}, one gets easily:
\begin{equation}
\label{quantqp} A_{ \bp }= \bP  \,,\quad
A_{q^{\beta}}=\frac{c_{\beta}}{c_{0}} \,Q^{\beta}\,,A_{ \bq }=\frac{\;c_{210}}{c_{0}} \bQ .
\end{equation}
Whereas $ \bQ $ is essentially self-adjoint, we recall that the operator $ \bP  $ is symmetric but has no
self-adjoint extension.  

The quantization of the kinetic energy gives
\begin{equation}
\label{qkinener} A_{p^{2}}=P^{2}+KQ^{-2}\,, \quad  K=2\pi\langle P^{2}\psi|\psi\rangle\,,\quad
\end{equation}
 We see that a repulsive potential $\propto\, 1/Q^2$ is obtained with strength 
$K>0$. In a semi-classical interpretation, this extra centrifugal term prevents the particle to reach the singular point, whatever the value of its angular momentum \cite{gazkoimur17}.
With a suitable choice of $\psi$, 
we can make this coefficient large enough, namely $K \geq 1$ \cite{reedsimon2,kowal02,asch07}, 
to make \eqref{qkinener} an essentially self-adjoint kinetic operator, i.e. no boundary condition is needed,  and then quantum dynamics of the free motion on the punctured plane is unique. Whilst canonical quantization, based on Weyl-Heisenberg symmetry which is unnatural in the present case,   introduces ambiguity on the quantum level, ACS quantization with suitable fiducial vector removes this ambiguity.  

 The quantum states and their dynamics have semi-classical phase space
representations through symbols. For the state
$|\phi\rangle$ the corresponding symbol reads
\begin{equation}
\label{Phisym} \Phi( \bq , \bp )=\frac{\langle  \bq , \bp |\phi\rangle}{(2\pi)^2}\,,
\end{equation}
with the associated probability distribution on phase space given
by
\begin{equation}
\label{rhophi} \Upsilon_{\phi}( \bq , \bp )=\dfrac{1}{2\pi c_{0}}|\langle
 \bq , \bp |\phi\rangle|^{2}.
\end{equation}
Having the (energy) eigenstates of some quantum Hamiltonian $\mathsf{H}$ at
our disposal, the most natural one being in this context the quantized $A_h$ of a classical Hamiltonian $h( \bq , \bp )$, we can compute the time evolution
\begin{equation}
\label{rhophiev} \Upsilon_{\phi}( \bq , \bp ,t):=\dfrac{1}{2\pi
c_{0}}|\langle  \bq , \bp |e^{-\ii \mathsf{H}t}|\phi\rangle|^{2} \, , 
\end{equation}
for any state $\phi$. 

The map \eqref{lowsymb}, yielding lower
symbols from classical $f$ reads in the present case:
\begin{equation}
\label{afflowsymb} 
\begin{split}
\widecheck{f}( \bq , \bp )
&=
\frac{(2\pi)^2}{c_0}\int_{\R_{\ast}^2}\frac{\ud^2
 \bq^{\prime}}{q^2\;{q^{\prime}}^2}\,\int_{\R_{\ast}^2}\ud^2  \bx  
\,\int_{\R_{\ast}^2}\ud^2
\by \,e^{\ii  \bp (\by- \bx  )}\hat{f}_p( \bq^{\prime},\by- \bx  )\times\\
&\times \psi\left(\frac{ \bx  }{ \bq }\right)\,
\psi\left(\frac{ \bx  }{ \bq^{\prime}}\right)\,\psi\left(\frac{y}{ \bq }\right)\,
\psi\left(\frac{\by}{ \bq^{\prime}}\right)\,,
\end{split}
\end{equation}
where $\hat{f}_p$ stands for the partial inverse Fourier transform introduced in \eqref{parcoure}, and with implicit hypotheses on $f$ and $\psi$ allowing derivations,  Fourier transform and permutation of integrals allowed by the Fubini theorem.

For functions $f$ depending on $ \bq $ only, expression
\eqref{afflowsymb} simplifies to a lower symbol  depending on $ \bq $
only:
\begin{equation}
\label{lowfq} \widecheck{f}( \bq )=
\frac{(2\pi)^2}{c_0}\int_{\R_{\ast}^2} \frac{\ud^2
 \bq^{\prime}}{q^2\;{q^{\prime}}^2}\, f( \bq^{\prime})  \int_{{R_{\ast}^2}}\ud^2
 \bx  \,\psi^2\left(\frac{ \bx  }{ \bq }\right)\,\psi^2\left(\frac{ \bx  }{ \bq^{\prime}}\right)\,.
\end{equation}
For instance, any power of $q= \sqrt{q_1^2 + q_2^2}$ is transformed into the same power
up to a constant factor
\begin{equation}
\label{powq} q^{\beta} \mapsto  \widecheck{q^{\beta}}=\frac{c_{\beta}c_{-\beta-2}}{c_{0}} \, q^{\beta}\,.
\end{equation}
For $f( \bq , \bp )= \bp $, we prove that:
\begin{equation}
\label{lsymvp}
\widecheck{ \bp }= \bp .
\end{equation}
For $f( \bq , \bp )= p^2$, we have:
\begin{equation}
\widecheck{p^2}=p^2+\frac{\gamma^2}{q^2}\, , \quad \gamma^{2}=2\left\langle P^2\psi|\psi\right\rangle\,. 
\end{equation}
Another interesting formula in the  semi-classical context
concerns the Fubini-Study metric derived from the symbol of total
differential $\ud$ with respect to parameters $ \bq $ and $ \bp $ affine
coherent states,
\begin{equation}
\ud\sigma( \bq , \bp )^2=\Vert\ud| \bq , \bp \rangle\Vert^2-\vert\langle  \bq , \bp |\ud| \bq , \bp  \rangle\vert^2
\end{equation}
Choosing a $\psi$ such that $c_{-2 0 1}=0$ 
gives:
\begin{equation}
\label{dercs} \vert\lg  \bq , \bp |\ud| \bq , \bp \rg\vert^2=c_{-2 1 0}^2[q_1^2 \ud p_1^2+q_2^2 \ud p_2^2+2q_1 q_2 \ud p_1 \ud p_2]\, . 
\end{equation}

For the   squared norm of $\ud| \bq , \bp \rg$,
\begin{align}
\ud\sigma( \bq , \bp )^2&= q^{-4}(A^2 q_1^2+B^2 q_2^2) \ud q_1^2+q^{-4}(B^2q_1^2+A^2 q_2^2) \ud q_2^2 +\\\nonumber
&+ 2q^{-4}((A^2-B^2)q_1 q_2+C(q_1^2-q_2^2) )\ud q_1 \ud q_2 +\\\nonumber
&+((E^2 -c_{-201}^2) q_1^2+F^2 q_2^2) \ud p_1^2+(F^2q_1^2+(E^2-c_{-201}^2) q_2^2) \ud p_2^2 +\\
\nonumber &+ 2((E^2-F^2-c_{-201}^2)q_1 q_2+G(q_1^2-q_2^2) )\ud p_1 \ud p_2\, , 
\end{align}
where:
\begin{align}
& A= \int_{\R_{\ast}^2}\frac{\ud^2\by}{y^2}\,\left(\by\cdot \pmb{\nabla}_{\by}[y\psi(\by)]\right)^2 \, ,  \\\nonumber
& B= \int_{\R_{\ast}^2}\frac{\ud^2\by}{y^2}\,\left(\by\times \pmb{\nabla}_{\by}[y\psi(\by)]\right)^2\, ,\\\nonumber
& C= \int_{\R_{\ast}^2}\frac{\ud^2\by}{y^2}\,\left(\by\cdot \pmb{\nabla}_{\by}[y\psi(\by)]\right) \,\left(\by\times \pmb{\nabla}_{\by}[y\psi(\by)]\right)\, ,\\\nonumber
& D= \int_{\R_{\ast}^2} \ud^2 \by (\psi(\by))^2 y_1^2\, ,\\\nonumber
& E=  \int_{\R_{\ast}^2} \ud^2 \by (\psi(\by))^2 y_2^2\, ,\\\nonumber
& F= \int_{\R_{\ast}^2} \ud^2 \by (\psi(\by))^2 y_1 y_2\, .\\\nonumber
\end{align}

\subsection*{Acknowledgements}
J.-P. Gazeau thanks  the ICTP Trieste for financial support and  hospitality. 
T.Koide acknowledges the financial support by CNPq (303468/2018-1) and  
a part of the work was developed under the project INCT-FNA Proc.\ No.\ 464898/2014-5.

%
 \section*{Conflict of interest}

 The authors declare that they have no conflict of interest. \\
 The ideas and opinions expressed in this article are those of the authors and do not  necessary represent the view of UNESCO.



\end{document}


\begin{abstract}
A few errors have been identified in  the article [Adv. Oper. Theory (2020) 5:901-935] and they are  corrected in this Addendum. Furthermore,  some  notations have been modified in order to avoid any confusion. 
\end{abstract}


\maketitle
PACS: {81R05,  81R30,  81S05, 81S10,  81S30,  81Q20}
\tableofcontents

\section{Introduction}
A set of errors have been identified in  the article \cite{gakomu20} and they are corrected in this erratum. Furthermore,  some  notations have been modified (and simplified) in order to avoid any confusion. 
\section {About ``Vectors as complex numbers'' in Subsection 3.1}
\label{noti}
In the paragraph 
\subsubsection*{Vectors as complex numbers}
the sentence 
\begin{quote}
\textit{In the sequel, we consider vectors $ \bq $ and $ \bp $ as biunivocally associated with the complex numbers $z( \bq )= qe^{\ii \theta}\equiv  \bq $ and $z( \bp )\equiv  \bp $ (same notation) respectively, and $\mathbf{1}\equiv 1$. }
\end{quote}
should be modified and completed as 
\begin{quote}
\textit{In the sequel, we consider  vectors $ \bq=(q_1,q_2) $ and $ \bp= (p_1,p_2) $, with  coordinates $q_i$, $p_i$, $i=1,2$,  with respect to an orthonormal basis $\{\mathbf{e}_1, \mathbf{e}_2\}$, as biunivocally associated with complex numbers through the multiplication rule 
$\mathbf{q}\mathbf{ q}^{\prime}= (q_1q^{\prime}_1-q_2q^{\prime}_2,q_1q^{\prime}_2 +q_2q^{\prime}_1)$. Thus $\mathbf{e}_1 \mathbf{q} = \mathbf{q}$ for all $\mathbf{q}$ whereas $\mathbf{e}_2 \mathbf{e}_2=-\mathbf{e}_1$, so $\mathbf{e}_2(q_1,q_2)= (-q_2,q_1)$.   Within the context of this paper, one thus avoids the confusion  between this ``imaginary'' $\mathbf{e}_2$ and  the ``$\ii$'' appearing in canonical quantization $ \bp \mapsto \mathbf{ P}=-\ii \pmb{\nabla}$, where $\pmb{\nabla}=\left(\dfrac{\partial}{\partial q_1},\dfrac{\partial}{\partial q_2}\right)$. 
The Euclidean inner product is indifferently denoted $\mathrm{Re}(\bq^\ast \bq')\equiv \bq\cdot\bq^{\prime}$. Consistently, $\mathbf{e}_1\cdot \bq=q_1$ and $\mathbf{e}_2\cdot \bq=q_2$. 
Also note that we are indifferently using $\mathbf{e}_1=(1,0)$ or $1$ when there is no risk of confusion. For instance, we will  occasionally write $\left.F(\bq)\right\vert_{\bq=\mathbf{e}_1}= F(1)$.}
\end{quote}

\section {Modification of notations and corrections in Subsubsection 6.2.2 ``Separable functions $f$''}
\label{sss622}
\begin{enumerate}
  \item After Eq. (6.23) the following should be added: 
 \begin{quote}
\textit{Here $(\pmb{\nabla} \widehat{\vap}_p)(1,-\cdot):= \left[\left(\dfrac{\partial}{\partial q_1}\widehat{\vap}_p)(\bq,-\cdot),\dfrac{\partial}{\partial q_2}\widehat{\vap}_p)(\bq,-\cdot\right)\right]_{ \bq= \mathbf{1} }$. More generally and from now on, the  gradient vector $ \pmb{\nabla}F(\bq)$ of a  function $F(\bq)$,  is as well  viewed as the complex number   $\left(\dfrac{\partial}{\partial q_1}F(\bq),\dfrac{\partial}{\partial q_2}F(\bq)\right)$
 associated with $\{\mathbf{e}_1, \mathbf{e}_2\}$.}\end{quote}
  \item Consistently with the above, Eqs. (6.24) and (6.25) can be replaced by the simpler one:
  \begin{equation*}
A^{\vap}_{ \bp } =  \bP  + \frac{\ii}{\bQ^{\ast}}\left(2+\frac{\pmb{\nabla}\Omega_0(1)}{\Omega_0(1)}\right)\, , 
\end{equation*}
  \item Similarly, Eqs. (6.26) and (6.27) should, after correction, be replaced by the simpler one:
  \begin{equation*}
  A^{\vap}_{ \bp  ^{\,2}}= \bP^2+\frac{2\ii}{Q^2}\left[2\bQ+\frac{\bQ\pmb{\nabla}\Omega_{0}(1)}{\Omega_{0}(1)}\right]\cdot  \bP   -\frac{1}{Q^2}\left[4+4\frac{\mathbf{e}_1\cdot\pmb{\nabla}\Omega_{0}(1)}{\Omega_{0}(1)}+\frac{\triangle\Omega_{0}(1)}{\Omega_{0}(1)}\right]\,,
 \end{equation*}
 \item Eqs. (6.28) and (6.29) should be replaced by:
 \begin{equation*}
A^{\vap}_{ \bq \cdot  \bp }=\frac{2\pi}{c_{\sfMv}}\left({\Omega}_{(2,1,0)}(1)( \bQ \cdot \bP   +2\ii)-\Omega_{(2,0,1)}  \bQ \times \bP  +\ii(\mathbf{e}_1\cdot\pmb{\nabla}\Omega_{(2,1,0)}(1)-\mathbf{ e}_2\cdot\pmb{\nabla}\Omega_{(2,0,1)}(1)) \right).
\end{equation*}
 \item Finally, Eqs. (6.30)  should be replaced by:
 \begin{equation*}
A^{\vap}_{ \bq \times  \bp }= 
\frac{2\pi}{c_{\sfMv}}\left(\Omega_{(2,1,0)}(1) \bQ \times  \bP  +\Omega_{(2,0,1)}(1)( \bQ \cdot \bP  +2\ii)
+ \ii(\mathbf{ e}_1\cdot\pmb{\nabla}\Omega_{(2,0,1)}(1)+\mathbf{ e}_2\cdot\pmb{\nabla}\Omega_{(2,1,0)}(1))\right)\,.
\end{equation*}
\end{enumerate}

\section {Other corrections}
\begin{enumerate}
  \item In the first formula of the proof of Proposition 1, the two last lines should read
 \begin{align*}
&= \int_{\mathbb{R}_{\ast}^{2}}\ud^2  \bx   \,\overline{\phi_1( \bx  )}\left\{\int_{\mathbb{R}_{\ast}^{2}} \ud^2  \bx^{\prime} \,,
\mathcal{M}^{\vap}( \bx  , \bx^{\prime})\phi_2( \bx^{\prime})\right\}\,,\\ 
&\mathcal{M}^{\vap}( \bx  , \bx^{\prime})= \frac{1}{2\pi} \frac{x^2}{{x^{\prime}}^2}\, \widehat{\vap}_p\left(\frac{ \bx  }{ \bx^{\prime}},- \bx  \right)\,,
\end{align*}  
  \item Equation (4.14), namely  $\mathrm{Tr}\left\{U( \bq , \bp ) \, C_{DM}^{-1}\right\}=\frac{1}{q} \mathrm{Tr}\left\{C_{DM}^{-1} \,U( \bq , \bp ) \right\}$, should be replaced with
  \begin{equation*}
U( \bq , \bp ) \, C_{DM}^{-1}=\frac{1}{q} \,C_{DM}^{-1} \,U( \bq , \bp ) \,.
\end{equation*}
  \item Section 5, 2d-3d line, `\textit{`inversion of operator}'' should read ``\textit{inversion  operator}''.
\item Below Eq. (6.15), replace  $w( \bx  )=\widehat{\vap}_p(1,- \bx  )$ with $w( \bx  )=2\pi\,\widehat{\vap}_p(1,- \bx  )$

\item Replace Eq. (6.18) (where parentheses where missing in subscripts) by
\begin{equation*}
\label{Omuqq}
\Omega_{(\beta, \nu_1,\nu_2)}( \bu):= \int_{\mathbb{R}_{\ast}^2}\frac{\ud^2 \by}{y^{\beta +2}}\,\widehat{\vap}_p\left( \bu,-\by\right)\,y_1^{\nu_1}\,y_2^{\nu_2}\,, \quad \Omega_{(\beta,0,0)}( \bu)\equiv \Omega_{\beta}( \bu)\,,
\end{equation*}
\item In Eq. (6.22) the binomial coefficient $\binom{s}{n}$ is wrong and should be replaced with $\binom{n}{s}$
\item Replace  Eq. (B2) with the following one:
\begin{equation*}
\label{Ompsi1a} 
(\pmb{\nabla} \Om)(1)= -2\Omega(1)\mathbf{ e}_1-\\i2\pi\left\langle \frac{1 }{ \bQ } \bP \psi|\psi \right\rangle\, .
\end{equation*}
\item Replace  Eq. (B3) with the following one:
\begin{equation*}
\label{Ompsi1b} 
(\Delta \Omega)(1))=4\Omega(1)+8\ii\pi \left\langle \mathbf{ e}_1\cdot\left(\frac{1}{ \bQ } \bP\right)\psi|\psi\right\rangle-2\pi\left\langle  \bP  ^{2}\psi|\psi\right\rangle\, .
\end{equation*} 
\item Replace  Eq. (B4) with the following one:
\begin{equation*}\label{psireal1} 
2\mathbf{ e}_1 +\frac{\pmb{\Omega}^{(1)}(1)}{\Omega(1)}=\mathbf{0}\,.
\end{equation*}
\item Replace  Eq. (B5) with the following one:
\begin{equation*}\label{psireal2} 
4+4\frac{\mathbf{ e}_1\cdot\pmb{\Omega}^{(1)}(1)}{\Omega(1)}+\frac{\Omega^{(2)}(1)}{\Omega(1)}=
-2\pi\langle  \bP  ^{2}\psi|\psi\rangle\,. 
\end{equation*}
\item In the proof below Eq. (B5) replace $\frac{\bP }{ \bQ}$ with $\frac{1 }{\bQ } \bP$. 
\end{enumerate}